\documentclass{aa}

\usepackage[normalem]{ulem}
\usepackage{graphicx}
\usepackage{txfonts}
\usepackage{pdfpages}
\begin{document}

   \title{MPS-ATLAS: A fast all-in-one code for synthesising stellar spectra}


   \author{V. Witzke\inst{1}\fnmsep\thanks{e-mail: witzke@mps.mpg.de}
          \and
          A.~I.~Shapiro\inst{1}
          \and 
          M.~Cernetic\inst{1}
          \and
          R.~V.~Tagirov\inst{1,2}
          \and
          N.~M.~Kostogryz\inst{1}
          \and
          L.~S.~Anusha\inst{1}
          \and
          Y.~C.~Unruh\inst{2}
          \and
          S.~K.~Solanki\inst{1,3}
          \and
          R.~L.~Kurucz\inst{4}
          }

   \institute{Max Planck Institute for Solar System Research, Justus-von-Liebig-Weg 3, 37077 G\"ottingen, Germany\\
         \and
         Blackett Laboratory, Imperial College London, South Kensington Campus, London SW7 2AZ, UK. \\
         \and
         School of Space Research, Kyung Hee University, Yongin, Gyeonggi, 446-701, Republic of Korea\\
         \and
         Center for Astrophysics | Harvard \& Smithsonian , 60 Garden Street, Cambridge, MA 02138, USA  \\
             }

   \date{---}

  \abstract
   {Stellar spectral synthesis is essential for various applications, ranging from determining stellar parameters to comprehensive stellar variability calculations. New observational resources as well as advanced stellar atmosphere modelling, taking three dimensional (3D) effects from radiative magnetohydrodynamics calculations into account, require a more efficient radiative transfer. }
   {For accurate, fast and flexible calculations of opacity distribution functions (ODFs), stellar atmospheres and stellar spectra we developed an efficient code building on the well-established ATLAS9 code. 
   The new code also paves the way for an easy and fast access to different elemental compositions in stellar calculations.
   }
   {For the generation of ODF tables we further developed the well-established DFSYNTHE code by  implementing additional functionality, and a speed-up by employing a parallel computation scheme. In addition, the line lists used can be changed from  Kurucz's recent lists. In particular, we implemented the VALD3 line list. 
   }
   {A new code, the \textbf{M}erged \textbf{P}arallelised \textbf{S}implified ATLAS is presented.   It combines the efficient generation of ODF, atmosphere modelling and spectral synthesis in  local thermodynamic equilibrium, therefore being an all-in-one code.  This all-in-one code provides more numerical functionality and is substantially faster compared to other available codes. The fully portable MPS-ATLAS code is validated against  previous ATLAS9 calculations, the PHOENIX code calculations, and high quality observations.}
   {}


   \maketitle
%

\section{Introduction}
Electromagnetic radiation emitted from the stellar photosphere is one of the key sources of information about a star. To advance our understanding of stars, accurate spectroscopic and photometric measurements, as well as accurate modelling that allows us to connect the measured stellar electromagnetic radiation to properties of the stellar atmospheres and stellar interiors are essential.

The data gathered by the  numerous ground-based and space-borne  telescopes that started observing during the last decade have highlighted the need for accurate modelling of stellar atmospheres and their spectra.   For example, the Echelle Spectrograph for Rocky Exoplanet- and Stable Spectroscopic Observations \citep[ESPRESSO, see][]{ESPRESSO} and the High Accuracy Radial Velocity Planet Searcher \citep[HARPS, see][]{HARPS} made high resolution spectral data  for thousands of stars available, while the Large Sky Area Multi-Object Fibre Spectroscopic Telescope  \citep[LAMOST, see ][]{LAMOST2016} provides low-resolution spectra for millions of stars. The interpretation of these spectral measurements requires modelling of stellar atmospheres on a fine grid of stellar fundamental parameters, such as effective temperature, surface gravity, and chemical composition.

Furthermore, the advent of planetary hunting missions, e.g. \textit{Kepler} \citep{KEPLER}, TESS \citep{TESS}, CHEOPS \citep{2020CHEOPS}, and WASP \citep{2006WASP} brought measurements of the photometric variability for several hundred thousands of stars. Even more data are expected from the forthcoming PLATO mission \citep{PLATO}. The main source for photometric variability of cool stars, like the Sun, are surface magnetic fields that affect the local structure in stellar atmospheres. Consequently, interpreting stellar photometric data requires an assessment of the effect of magnetic field in stellar atmospheres on the emergent radiation.

There are different approaches to modelling stellar spectral and photometric fluxes. One of the simplest, and most widely used approach relies on 1D modelling of stellar atmospheres under the assumption of radiative-convective equilibrium (with a simple parameterisation for convective flux and overshooting). While such a 1D approach has a number of shortcomings \citep[see, e.g.][]{koesterkeetal2008, 1D_bad} it proved itself to be an invaluable tool for various applications \citep[see, e.g.][]{1994A&A...281..817C, Claret_CLV_2000AA, Apogee_1D_models_2012, Carmens_stellar_parameters_2020} and is extensively used in stellar physics.

A more comprehensive approach relies on 3D hydrodynamic and magnetohydrodynamic (HD and MHD, respectively) simulations of near-surface convection in stars \citep[see, e.g.][]{ nordlund_lr_2009, stein_lr_2012, Freytag_2012JCoPh,  Magic_2013A&A_stagger01, beeck_i_2015, beeck_ii_2015}. Using the simulated 3D cubes the emitted radiation can be calculated following a 1.5D approach, i.e. along many rays passing through such a 3D cube \citep[see,][for a detailed description of the 1.5D approach]{2006Asplund_3D_1Dcomp, rietmueller-solanki-2014, norris-beck-2017}.
An important advantage of the 3D MHD simulations and 1.5D approach over 1D radiative equilibrium modelling is that it allows us to directly account for the effects of the magnetic field on the emergent radiation and, consequently, to model the stellar spectral and photometric variability.

Altogether, there are two separate but similar challenges that demand  comprehensive and fast spectral synthesis on an adjustable frequency grid, and broad band spectral intervals. First, the 1.5D modelling needs a huge amount of fast radiative transfer (RT) computations on 1D structures. Second, accurate and fast 1D atmosphere modelling with subsequent RT calculations for any stellar parameter is needed. 
The aim of the present study is to develop a fast and easily applicable RT code to address both these challenges.

The  accurate treatment of the line opacity poses the main challenge to spectral synthesis over broad spectral ranges because of the immense number of spectral lines:  where current line lists contain up to hundreds of millions of  lines \citep{kurucz2005}  that need to be taking into account. A careful treatment of both atomic and molecular lines is imperative because atomic and molecular lines are interspersed in particular in the  spectra of cool stars. Spectral lines  not only dominate some spectral regions (e.g. the UV), but also affect the atmospheric structure by blocking photons. 
While a forward spectral synthesis on a high-resolution wavelength grid, i.e. typically with a resolving  power, $\rm R = 500 000$, is computationally expensive, not all applications require high resolution spectra. To that end, different techniques were developed to correctly account for line opacity, but reduce the computational cost on coarse resolution grids. The most commonly used methods are the opacity distribution functions (ODF) method and the Opacity Sampling (OS) method \citep{Carbon_book_1984, Castelli_ATLAS12_2005}. Both methods approximate the opacity using different ways of sampling it. We preferred the ODF method for developing the RT code, as it results in significantly faster RT calculations compared to the OS method,  thus making it more suitable for the 1.5D approach. 
Recently, the ODF approach was further optimised to make it more efficient. \citet{Cernetic_sub} found the best configurations of the ODF to reach several times faster computations while maintaining accuracy. Moreover, the ODF approach was extended to calculate stellar fluxes as they are observed in various filters  \citep{Cernetic_sub, Anusha_2020}.
%

Widely used RT codes used for spectral synthesis of cool stars include the local thermodynamic equilibrium (LTE) codes MARCS \citep{MARCS_descript2008}, and MAFAGS-OS \citep{Grupp_2004A&A},  and the non-LTE codes PHOENIX \citep{Phoenix_descript2013}, and TLUSTY \citep{Hubeny_2017_tlusty}. 
The PHOENIX code was developed to account for expanding atmospheres and deviations from LTE, and is therefore more complex and CPU intensive. 
Another successful RT code in LTE, which can calculate both model atmospheres in radiative equilibrium and the emergent spectra, is the ATLAS code by Kurucz \citep{Kurucz_manual_1970}. 
Since for most applications mentioned above the spectral range does not include the extreme and far UV, it is sufficient to consider LTE codes. Thus, the ATLAS code includes all crucial physics for radiative transfer in the atmospheres of main-sequence stars while keeping the setup simple. Consequently, we chose to build on the ATLAS code.  
While the ATLAS code comes in two versions; ATLAS12 \citep{Castelli_ATLAS12_2005} and ATLAS9, we prefer the ATLAS9 version as it uses ODF whereas the ATLAS12 works with the OS method.    
The advantage of the ATLAS9 code is the short time required to compute a single model (a few minutes on a single core using the standard ODF table with 328 bins).

When updating ATLAS9 it is essential to also consider the DFSYNTHE code \citep{Kurucz_2005_Atlas12_9, Castelli_2005_DFSYNTHE}. This code computes ODF tables for the ATLAS9 code.  The main disadvantage of the ODF so far was the limitation to the chemical composition and microturbulent velocities for which the ODF tables  were pre-tabulated.  The generation of ODFs using the DFSYNTHE code was not suitable for massive computations as several routines had to be successively executed, and the computation time was very long \citep{Kurucz_2005_Atlas12_9, Castelli_2005_DFSYNTHE}. 
To achieve our goal of fast spectral synthesis for arbitrary abundances, we need to eliminate this bottleneck, and in addition make the ODF generation more user friendly.
This is achieved by merging the DFSYNTHE code, which calculates high-resolution opacities and uses them to obtain ODF, with the ATLAS9 code that can calculate both the atmospheric structures as well as the emergent spectra. Additionally,  a more flexible treatment of the ODF was implemented. The frequency resolution, and ODF configuration can be changed, and a spectral filter included.  Furthermore, we parallelise the code to speed up the high resolution opacity calculations and model calculations.  Finally, the line lists used can be exchanged and we give example calculations with both, Kurucz's line list as well as the most up to date VALD3 line list \citep{VALD_2015}. In this paper we describe the structure, and improvements of the  resulting code, which we call the  \textbf{M}erged \textbf{P}arallelised \textbf{S}implified ATLAS code (MPS-ATLAS).  Moreover, we validate  MPS-ATLAS against previous ATLAS9 calculations,  PHOENIX code calculations, and observations.  The MPS-ATLAS code is available on request, and it will be made publicly available soon. 

The paper is structured as follows. A brief summary of the physics and its implementation is given in Section~\ref{Sec:RT-general} and  Section~\ref{Sec:Code_description}, respectively. All code improvements are listed in Section~\ref{Sec:Code_optimisations}, where we also discuss limitations of the code. In Sect.~\ref{Sec:Code_testing} we comment on the code performance and test  the resulting emergent spectra by a code-to-code comparison and a code to observations comparison. The outcome of the work is summarised in Sect.~\ref{Sec:Discussion}, where also conclusions are drawn.

%

\section{Radiative transfer calculations}
\label{Sec:RT-general}

The energy transport in  stars defines the structure of their interior and atmosphere, as well as the emitted  electromagnetic radiation. Focusing on the upper layers around the optical surface of a star, the radiative transport of energy is dominant.  Thus, the imperative problem for modelling stellar atmospheres is solving the radiation transfer equation (RTE)
\begin{equation}
\label{eq:RTE_general}
 \rm \mu \frac{dI_{\lambda}}{d \tau_{\lambda} } =  I_{\lambda}- S_{\lambda},
\end{equation}
along a ray for a time-independent system, where the subscript $\lambda$ denotes the wavelength and implies that all quantities are monochromatic.  $\rm I_{\lambda}$ is the intensity,  $ \rm S_{\lambda}$ is the source function, and $\mu = \cos \, \theta$,  where $\theta$ is the angle between the viewing direction and the normal to the stellar surface. The optical depth $\rm \tau_{\lambda}$ is determined from \mbox{ $\rm d\tau_{\lambda}=-\chi_{\lambda}(s)\, ds$}, where $\rm s$ is the height in the atmosphere.  
The total extinction coefficient per unit volume is $\rm \chi_{\lambda} = \alpha_{\lambda} + \sigma_{\lambda}$, which contains the absorption coefficient, $\rm \alpha_{\lambda}$, and the scattering coefficient,  $\rm \sigma_{\lambda} $.    
The source function is generally defined as the ratio of the total emission coefficient, $\rm j_{\lambda}$, to the total extinction coefficient
\begin{equation}
  \rm   S_{\lambda} = \rm j_{\lambda} /\rm \chi_{\lambda}. 
\end{equation}

Solving the RTE becomes straightforward once the emission coefficient and opacity are known along the atmosphere. 
In  local thermodynamic equilibrium (LTE) and under the assumption of coherent isotropic scattering, the emissivity can be expressed as
\begin{equation}
\rm    j_{\lambda} = \alpha_{\lambda} B_{\lambda} + \sigma_{\lambda} J_{\lambda},
\end{equation}
where $\rm B_{\lambda}$ is the Planck-function,  and $\rm J_{\lambda}$ is the mean intensity \citep{FRH_Mihalas}.
Then, calculating the opacity becomes the keystone to solving the RTE (except scattering coefficients).

\subsection{Calculating opacity}
\label{subsection:calc_the_op}
%
The opacity $\kappa_{\lambda}$ (hereafter, we refer to the opacity normalised per unit of mass $\kappa_{\lambda} \equiv \rm  \chi_{\lambda} / \rho$, where $\rho$ is the density) can be decomposed into continuum opacity, $ \kappa_{\lambda, c}$, associated with different processes that involve atomic and molecular transitions with  non-discrete wavelength (i.e. bound-free and free-free transitions), and line opacity, $ \kappa_{\lambda,l}$, due to discrete transitions, i.e. atomic and molecular spectral lines. 
To determine the total opacity $\kappa_{\lambda} = \kappa_{\lambda, c} + \kappa_{\lambda, l}$, 
at each point, the atomic and molecular level populations have to be determined. 
Under the assumption of LTE, the atomic and molecular level populations in equilibrium can be calculated using the Saha-Boltzmann (SB) equation.  This implies that they depend only on  elemental composition (i.e. abundances), local temperature and pressure. In addition,  in stellar atmospheres the  Doppler broadening of lines occurs due to turbulence, which in 1D models is taken into account using the micro-turbulence parameter, $\rm v_{turb}$.  Hence, for a given composition, opacity  is a function of wavelength, local temperature, pressure, and micro-turbulence  $\rm \kappa \equiv  \kappa (\lambda, T, P, v_{turb})$. 

In addition to the temperature and pressure values,  calculating the level populations using the SB equation requires knowledge of the electron number density, $n_e$, which however is not known a priori. Thus under the assumption of particle conservation the SB equation has to be solved iteratively. For a detailed description of the solver implemented in MPS-ATLAS  see Appendix \ref{app:solve_Enumbers} and \citet{Kurucz_manual_1970}. 

Having determined the populations, the continuous opacity, $\kappa_{\lambda, c}$, and the line opacity, $\kappa_{\lambda, l}$, can be calculated. For the continuous opacity, the MPS-ATLAS code takes  the following contributors into account:  Free-free (ff) and bound-free (bf) transitions in $\text{H}^-$, $\text{H}_2$, $\text{He}$, $\text{He}^-$, C, N, O, Ne, Mg, Al, Si, Ca, Fe, the molecules CH, OH and NH, and their ions. While for the calculation of the  equilibrium number densities and the line opacity also $\text{C}_2$ and CN are included, their contribution to the continuum opacity via photodissociation is neglected.  Moreover, electron scattering and Rayleigh scattering on $\text{H I}$, $\text{He I}$, and $\text{H}_2$ were considered. 

The line opacity is more expensive to compute, simply because of the huge number of lines for which  the line absorption coefficient needs to be calculated. For atomic and molecular transitions from an initial energy level, denoted by i to a final energy level j, the line absorption is defined as:
\begin{equation}
\label{eq:l_nu_line}
 \rm   \ell_{\nu} =  \frac{\sqrt{\pi}e^2}{m_e c} \frac{ g_i f_{ij} }{\rho \Delta {\nu}_D } \frac{\rm N_k }{\rm U_k }  e^{\frac{-E_i} {k_B T} } \left(1-e^{\frac{-h \nu_0}{ k_{B} T}} \right) H\left(\frac{\Delta {\nu}}{\Delta {\nu}_{D}}, \frac{\gamma}{4 \pi \Delta {\nu}_{D}} \right), 
\end{equation}
where $\rm \nu_0$ is the frequency corresponding to the transition, $\rm e$ it the elementary charge, $\rm m_e$ is the electron rest mass, $\rm c$ is the speed of light, $\rm g_{i}$  is the statistical weight of the level i, $\rm f_{ij}$ is the dimensionless oscillator strength of the transition $\rm i \xrightarrow{} j $,   and  $\Delta {\nu}_D$ is the  Doppler width. The gas density is $\rm \rho$, the temperature is $\rm T$, and $E_i$ is the energy of the initial energy level. The number density over partition functions for the entire ionisation stage, is given by  $\rm N_k / U_k$, where the subscript $\rm k$ indicates the ionisation stage. The term in brackets in Equation \eqref{eq:l_nu_line} describes the correction for stimulated emission, where the Boltzmann constant, $\rm k_{B}$, and the Planck constant, $\rm h$, have their usual designations. 
For the line broadening the Voigt function $\rm H(\Delta {\nu} / \Delta {\nu}_{D}, \gamma/ 4 \pi \Delta {\nu}_{D}  )$, where $\rm \gamma$ is the total damping constant and $ \rm \Delta {\nu} = {\nu} - {\nu}_{0} $, is used.  While for most lines the Voigt profile is used, for hydrogen lines more accurate profiles are employed, i.e.~Stark profiles are taken  into account \citep{SYNTHE_2002_Cowley}.

\subsection{Atmosphere models in radiative equilibrium and emergent spectra} 
\label{section:main:model_emergent_spec}
Modelling a stellar atmosphere for a particular set of stellar fundamental parameters and elemental composition  involves an iterative process of recalculating the atmospheric structure until radiative equilibrium (RE) is reached \citep[see for example][and references therein]{1989Collins_textbook, hubeny_mihalas_2015}. In this process, in each iteration the RTE for a currently estimated atmospheric structure has to be solved in order to determine by how much the model structure needs to be corrected to satisfy the RE (for convergence criteria see Appendix~\ref{app:Improv_model}). The iterations are usually initialised  with a starting structure that represents a converged solution for some other set of stellar fundamental parameters and composition. Before the iterative procedure starts, using the effective temperature of this starting structure,  the temperature value at each of its depth points is re-scaled by applying  the ratio of this effective temperature to the required one (for more details see Appendix~\ref{app:Re-calc}). 

Then the RE calculations start. At each of the iterations, the wavelength- and depth- dependent source function (consisting of the thermal and scattering parts) needs to be determined. 
This can be achieved by various methods. By default MPS-ATLAS uses a Feautrier method, but a second iterative method is also implemented (more details see Appendix \ref{app:RT-calc}).  Having found the source function, the moments of the intensity are calculated, namely the mean intensity $\rm J_{\lambda}$, and the flux, $\rm H_{\lambda}$.   This process is repeated for each wavelength point, typically in an interval from 9 nm to 10 000 nm, in order to obtain the total wavelength integrated Eddington flux, $\mathcal{H}$.

The aim is to match the total Eddington flux, $\mathcal{H}$, to the flux that corresponds to the desired effective temperature. Considering the flux, $\mathcal{H}$, and keeping in mind that the majority of it comes from the region where $0.1 \le \tau \le 2.0$, the temperature corrections $\delta T$ can be found in the optically thick regions. For that a modified version of the Avrett-Krook procedure is applied to take deviations from the RE due to convection and overshooting into account (for more details see \citet{Avrett_Krook_1963, Kurucz_manual_1970}). In the optically thin regions, the temperature does not greatly affect the overall flux, whereas the derivative of it, $ dH_{\nu}/d\tau_{\nu} =0 $, is sensitive  to a temperature change.  The  boundary condition for the  temperature corrections $\delta T$  in the optically thin part of the atmosphere requires the flux derivative of the Eddington flux to be zero \citep[for a discussion see][]{1978stat.bookMihalas}. This is achieved using a $\Lambda$ correction method  \citep{Boehm_vitense_1964}.   Then, the part of the atmosphere where both approaches overlap is smoothed out to match.

After the atmospheric structure is either calculated by the method described above or taken from some other source, e.g. from a ray through a 3D cube,  and the source function is found, the emergent intensity can be calculated by evaluating the integral
\begin{equation}
\label{eq:Ilambda}
    I_{\lambda} = \int_{0}^{\infty} S_\lambda\, e^{-\tau_{\lambda}/\mu } \rm d\tau_{\lambda} /\mu. 
\end{equation}
The integral in Eq.~\eqref{eq:Ilambda} has to be computed for each wavelength separately to get the whole emergent spectrum. Note, that the framework described here is restricted to a plane parallel setup. Thus, when using a computed 1D model the emergent intensities for different view angles $\mu$ can be obtained by setting a set of $\mu$ angles in the code. To calculate spectra emerging from 3D cubes for different view angles a 3D cube is  rotated and for each view angle a different set of 1D rays is obtained for which the RT is solved along the ray, but keeping the view angle  $\mu=1$.

\subsection{The ODF approach}
\label{subs:ODF_approach}


The synthesis of the line opacity is computationally demanding due to the tremendous number of spectral lines to be taking into account. For example, the default MPS-ATLAS  line list contains more than 100 million atomic and molecular lines \citep{kurucz2005}.
These millions of lines lead to very  complex and rich  spectra for cool stars. Consequently, a very fine wavelength grid is needed to catch all the details in the spectra. For a number of applications such detailed spectra are not required,  and low resolution calculations are sufficient. However, lines still affect even low-resolution spectra. A straightforward way  to include the effect of lines on low-resolution spectra is to simply average high-resolution spectra. Let us consider a small wavelength interval (hereafter, bin), e.g. between 0.1 and 10 nm wide,  between $\lambda_i$ and $\lambda_{i+1}$. This bin represents one low resolution grid point,  in which the intensity, $I_{\rm bin}$, representing the whole bin, has to be calculated 
\begin{equation}
    I_{\rm bin} = \int_{\lambda_i}^{\lambda_{i+1}} I_{\lambda} d\lambda. 
\end{equation}
The simplest way of approximating $I_{\rm bin}$ is by obtaining $I_{\lambda}$ on a fine wavelength grid using a large number of points, N, and taking the sum
\begin{equation*}
  I_{\rm bin} \approx  I_N = \sum_{j=1}^{j=N} I_{\lambda_j} \cdot \Delta \lambda_j,  
\end{equation*}
where $\Delta \lambda_j$ is the $j-$th discretised wavelength step. While this method is straightforward, the main issue is that $I_{\lambda}$ has to be calculated $N$ times for an accurate approximation, where $N$ for example is of order $O(10^4)$ for a one nanometre interval in the UV, if the resolving power R is 500 000. An alternative method to avoid solving the RT many times is the ODF approach \cite[see e.g.][pp. 625--627]{hubeny_mihalas_2015}, which approximates the complex structure of the opacity.

\begin{figure}[t]
  \centering 
  {\includegraphics[width=1.0\linewidth]{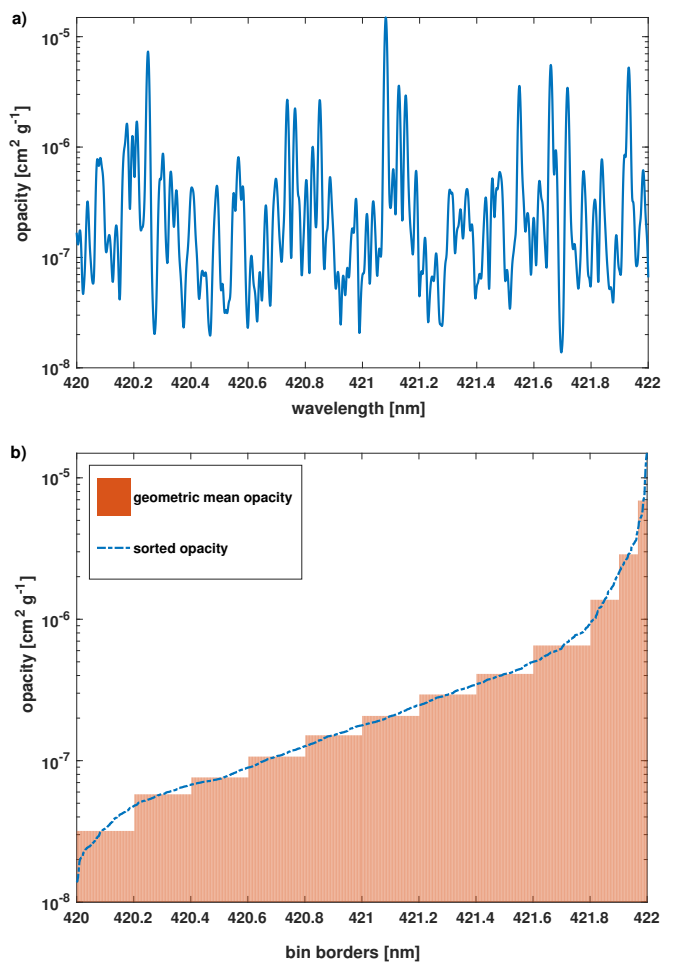}}
  \caption{Illustration of ODF generation in one example bin. a) detailed high-resolution opacity in the bin from 420 nm -- 422 nm. b) sorted opacity without the information of the wavelength, and the corresponding geometric mean values for 12 sub-bins, which were chosen as in \citet{Castelli_2005_DFSYNTHE}. }
  \label{fig:0003}
\end{figure}

The main goal of the ODF method is to reduce the number of points $N$ in a bin to a minimum, and still approximate the flux accurately.  
However, the opacity in one bin can abruptly change by several orders of magnitude within very narrow spectral intervals (see Fig.~\ref{fig:0003}a). Consequently, a fine spectral grid is required to accurately describe the opacity profile. One could think about averaging the opacity in the entire bin to avoid the high-resolution calculations.  In most cases such an averaging would lead to a gross overestimation of the line opacity effect, and would essentially lead to trapping photons that otherwise would escape \citep[see Fig.~2 and its detailed discussion in][]{Cernetic_sub}.

This can be circumvented by taking averages over wavelength points with similar opacity values, which can be achieved by grouping points in a certain opacity range together. A straightforward way is to sort the points in one bin in an ascending order by their opacity value. The sorted opacity profile can be described using significantly fewer points than that of the original opacity  (compare Figs.~\ref{fig:0003}a~and~\ref{fig:0003}b). The opacity profile in each bin is usually divided into several sub-bins, whereafter the opacity is averaged over each sub-bin using  the geometric mean (see Fig.~\ref{fig:0003} b). The geometric mean works better for the optically thick regime, because it avoids skewing the opacity towards large values. However, in the optically thin regime the arithmetic mean is better since the intensity depends on the opacity rather linearly. While both approaches are implemented, we used the geometric mean throughout this work in order to be consistent with the approach in the original version of the DFSYNTHE code.

Sub-bins might have  different widths, which are defined by the fraction of the whole bin size. Generally, the part with the greatest opacity values is subdivided into smaller sub-bins, while the part with smaller opacity values is split into a few large sub-bins.  The resulting step-function in the entire bin $[\lambda_{i}, \lambda_{i+1}] $, where each step is the averaged opacity $\kappa_{s,i}$ of the $s$-th sub-bin, is called opacity distribution function. 
For a more detailed description of ODF and  the importance of different sub-bin configurations, i.e. number of sub-bins and their widths, see \citet{Cernetic_sub} and references therein.

Then, the intensity in each of the sub-bins can be  calculated by solving the RT equation only once using the corresponding $\kappa_{s, i}$ value. For the entire $i$-th bin the intensity is obtained by summing over the contributions from the sub-bins, i.e. 
\begin{equation}
    I_N \approx I_{ODF} = \sum_{k=1}^{n_s} I_{s} \cdot \Delta \lambda_s,
\end{equation}
where $ I_s$ is calculated using $\kappa_{s, i}$, and $ \Delta \lambda_s$ is the sub-bin width.

Note that by re-arranging the opacity in a wavelength interval, as  is done during the sorting in the ODF approach, two important assumptions  are made implicitly: i)  the opacity shape in the bin does not change rapidly along the line of sight, in particular, within the region of the atmosphere where the radiation for the corresponding bin is formed \citep{1974Kurucz_ODFs} ii) the wavelength interval of the bins are small enough that changes of the Planck function as well as changes of the continuum opacity can be neglected.

As mentioned in Section~\ref{subsection:calc_the_op}  the opacity for a given elemental composition, and turbulent velocity, $\rm v_{turb}$ only depends on pressure and temperature  in the LTE case. Thus, the ODF table contains the information of the mean opacity in each sub-bin, for a given elemental composition and micro-turbulence and can be pre-tabulated on a T-P grid \cite[see e.g.][and references therein]{Castelli_2005_DFSYNTHE}. For more details on the implementation see Appendix~\ref{app:ODF-impl}. Having a pre-tabulated ODF table,   it is now straightforward for any stellar atmosphere to obtain the total opacity for any temperature and pressure using linear interpolation on the T-P grid and adding the continuous  opacity. 
Note, that when line opacity from the ODF tables is added to the continuous opacity the assumption that the continuous opacity in a particular bin is independent of wavelength  is made implicitly. This is currently a limiting factor for the upper limit of the bin size. One way to circumvent this limitation is to include the continuum opacity in the ODF tables. 

\section{Code structure}
\label{Sec:Code_description}
\begin{figure}
  \centering 
  {\includegraphics[width=1.0\linewidth]{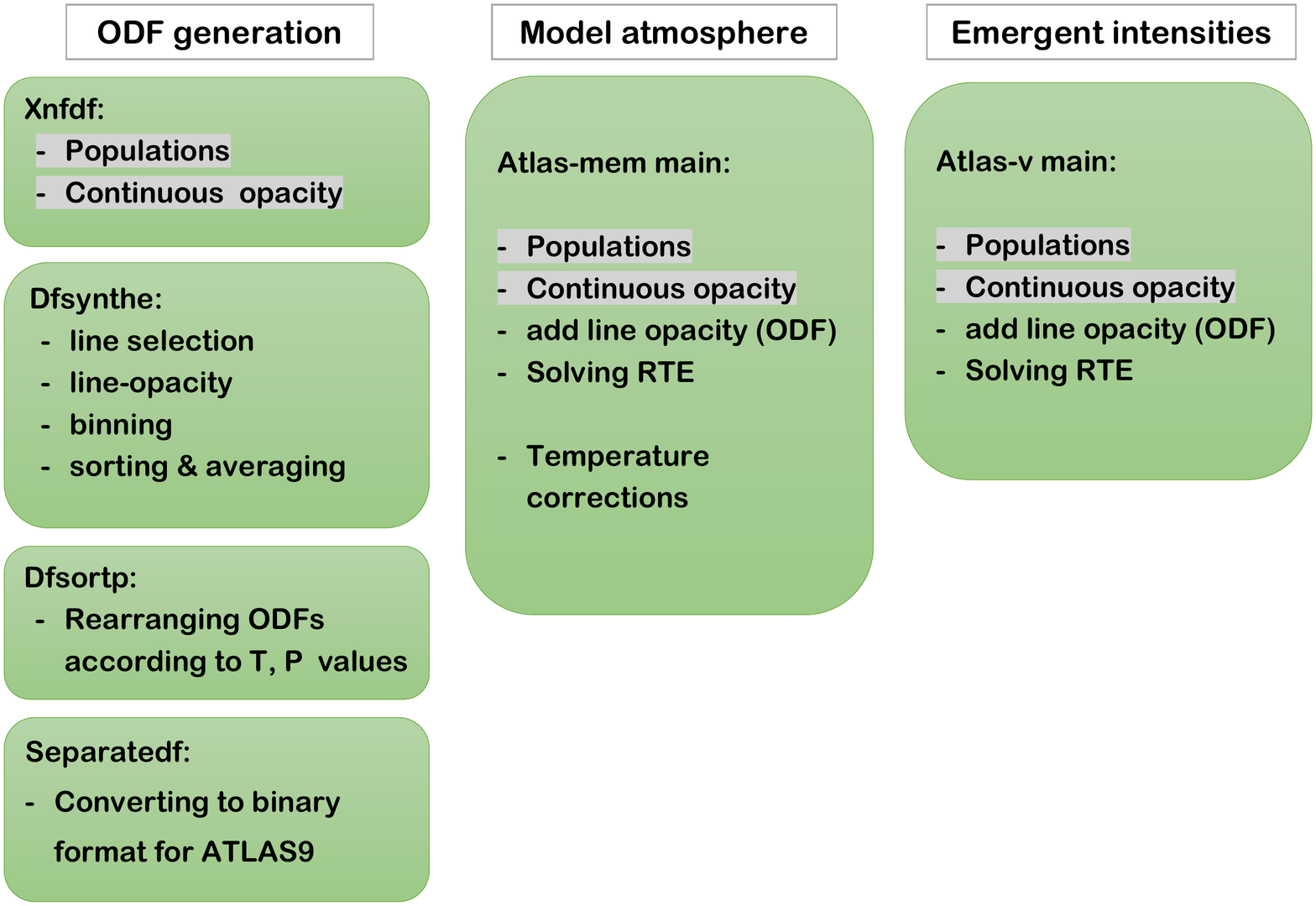}}
  \caption{Original structure of codes that calculate ODF,  model atmospheres, and emergent intensities. Each green bubble indicates a separate code that results in a separate executable. The listed tasks are the main procedures used in each program. Gray background indicates procedures that are the same for all three calculations. }
  \label{fig:0102}
\end{figure}
\begin{figure}
  \centering 
  {\includegraphics[width=1.0\linewidth]{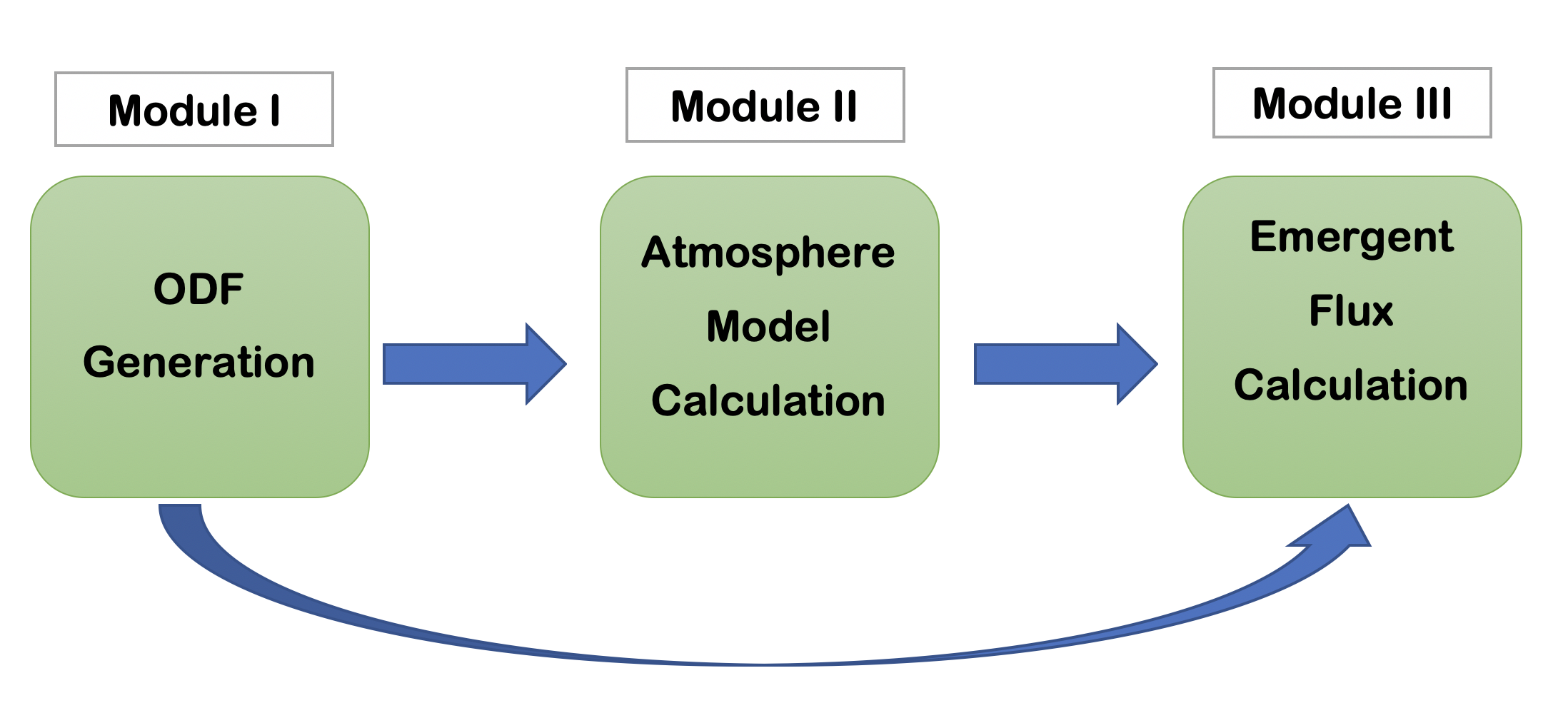}}
  \caption{ Schematic structure of the code modules I -- III. }
  \label{fig:0101}
\end{figure}
\begin{figure}
  \centering 
  {\includegraphics[width=1.0\linewidth]{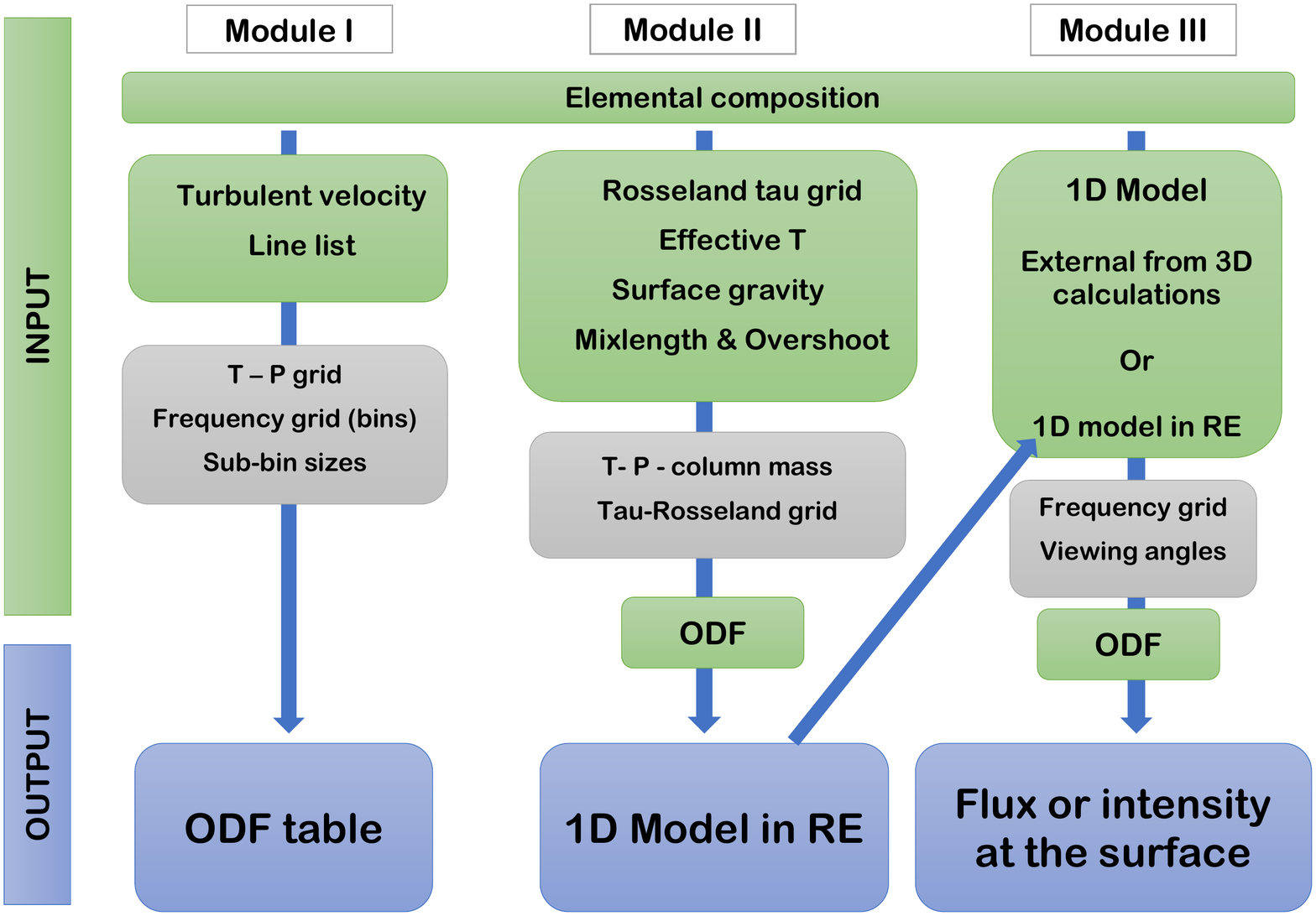}}
  \caption{Input parameters for different code modules together with information on the output.}
  \label{fig:0103}
\end{figure}

The MPS-ATLAS code results in one monolithic executable which depending on a control input file can perform a different set of calculations. 
An overview of the original three routines on which MPS-ATLAS is based, is given in Fig.~\ref{fig:0102}. Note, that the original ODF generation routine\footnote{http://wwwuser.oats.inaf.it/castelli/sources/dfsynthe.html} consists of four separate codes, resulting in four executable.
These three separate routines have been designed to perform the calculations outlined in Section~\ref{Sec:RT-general}  and correspond to the three MPS-ATLAS modules (see  Fig.~\ref{fig:0101} and Fig.~\ref{fig:0103}), where a module is a code internal execution mode.  The well-established DFSYNTHE routine \citep{Kurucz_2005_Atlas12_9, Castelli_2005_DFSYNTHE} corresponds to \mbox{module I} in the MPS-ATLAS code, and generates ODF tables (see left column in Fig.~\ref{fig:0102}).   Two slightly different ATLAS9 codes\footnote{https://wwwuser.oats.inaf.it/castelli/sources/atlas9codes.html} correspond to \mbox{module II} and \mbox{module III} in MPS-ATLAS, where one calculates 1D atmosphere models and the other the emergent intensity or flux (see middle and right columns in Fig.~\ref{fig:0102}). The main difference between the two ATLAS9 codes is the main routine. The code for calculating model atmospheres contains routines to perform the temperature corrections in addition to solving the RT.

The original collection of codes is intricate to run due to several codes that need consecutive execution. Furthermore, the three different codes make use of similar procedures. For example, both the DFSYNTHE and the ATLAS9 codes require calculations of the populations and  continuous opacity. While these calculations can be obtained using the same functions, both codes have their own copy of these functions with slightly different implementations. This renders any modification quite difficult.
Therefore, we chose to merge three codes into one code with different execution modes, that we call here modules,  that take care of the different calculations. The resulting, merged code is further adjusted, parallelised, and simplified, and thus called \textbf{M}erged-\textbf{P}arallelised-\textbf{S}implfied-ATLAS (MPS-ATLAS). MPS-ATLAS is mainly written in Fortran 90 (free form), but some parts are left in F77. We will rework them in future releases. We use dynamical memory allocation in several modules, for example in the ODF calculations.

The advantages of the MPS-ATLAS code are a more user friendly operation, significant computational speed-up, and wider functionality (in particular flexible ODF setup, which is discussed in Sec.~\ref{Sec:Code_optimisations}).   A schematic diagram in Fig.~\ref{fig:0101} shows the overall structure of the MPS-ATLAS code. The new code has three execution modes that correspond to the original structure of the three codes: ODF generation, model calculations, and emergent intensity or flux calculation.  These three modules can be either executed consecutively or separately, which is controlled during run-time. Both model calculations and emergent intensity calculations need ODF tables as input, which is  indicated by the arrow.  As discussed in Sect.~\ref{section:main:model_emergent_spec}, the model calculation is not necessary if alternative input atmospheres are used instead, for example a ray from a 3D radiative MHD cube (such a bypass of \mbox{module II} is indicated by the arrow in the bottom of Fig.~\ref{fig:0101}).

In the following we give an overview of the code structure in terms of input and output.   There are two types of parameters (shown in Figure~\ref{fig:0103}):  i) overall input parameters (highlighted in green), that are specified once and do not change for a particular calculation, for example the elemental composition ii) parameters that are set in the input, but which determine a grid on which the output calculations are performed (highlighted in grey). Figure~\ref{fig:0103} shows that the   calculations of an ODF table using \mbox{module I} requires the elemental composition, line lists, and the turbulent velocity. In addition, the temperature and pressure grid, as well as frequency range and the bin, sub-bin configuration  need to be specified (more details in Appendix~\ref{app:ODF-impl}). An example input file for the ODF calculation is shown in Appendix~\ref{app:example_input}. The output ODF is needed as input for both \mbox{module II}, and \mbox{module III}.

Module II calculates a 1D  atmosphere model in RE. As input parameters this module needs the elemental composition, effective temperature ($\rm T_{eff}$), surface gravity (log g), and the mixing-length parameter, which is needed to include convection through mixing length theory \citep{1958ZA.....46..108B}, for which overshoot can be turned on. In addition, an ODF table is required as input to account for the effect of line blanketing on the atmosphere structure. For the opacity table either the output from \mbox{module I}, or from a pre-calculated ODF table gird can be used. Moreover, an initial 1D model atmosphere is read, which is used as an initial starting point, and has the same format as the output model (for more details see Appendix~\ref{app:Re-calc}). The output model is obtained on an a priori specified $\tau_{\rm Ross}-$grid.
An atmosphere model consist of column mass, temperature, total gas pressure, electron number density ($\rm n_e$), Rosseland mean opacity ($\kappa_{\rm Ross}$), radiation pressure, and micro-turbulence velocity ($\rm v_{turb}$)  for each depth point. Currently, we only consider cases of height independent $\rm v_{turb}$, though for height-dependent micro-turbulence it is possible to use several ODF tables covering a range of $\rm v_{turb}$ values and interpolate between them.

The third module either synthesises the emergent flux or the emergent intensity for different view angles. Here, the only actual free parameters to set are the wavelength range and viewing angles, as the input is already pre-determined by the ODF and model atmosphere. While the elemental composition should match the ODF, the wavelength grid on which the ODF is calculated sets the wavelength grid of the emergent intensities. 
Moreover, the ATLAS output 1D atmosphere model in RE contains  all the necessary information to calculate the emergent spectrum, in particular the electron number density, and the radiative pressure at each depth point obtained with the elemental composition for which the model was calculated.  On the contrary,   if using an external model for example from a ray through a 3D MHD cube, several quantities might be unknown. In particular,   the electron number densities, which are needed to find the populations of all other ions, have to be obtained. In order to use external models that only have the column mass, pressure, and temperature,  an additional flag was introduced to recalculate the equilibrium number densities in each depth point for the given elemental composition. Example input files are discussed in more detail in Appendix~\ref{app:example_input}.


\section{Functionality extensions} 
\label{Sec:Code_optimisations}
In the following we describe the main functionality extensions included in the MPS-ATLAS code.

\subsection{Flexible wavelength grid and optimised sub-binning}

The DFSYNTHE code is limited to two particular wavelength-bin grids onto which ODFs are calculated, both with the same wavelength independent sub-bin sizes.   However, various applications require spectral synthesis with different spectral resolutions. Hence, it is important to have an option for changing the ODF wavelength grid. Furthermore, \cite{Cernetic_sub} showed that an optimal choice of the sub-bin grid leads to significant improvements in both accuracy and speed of the calculations. In particular, they showed that for most of the wavelength range two to four sub-bins are sufficient to reach the same accuracy as the standard 12 sub-bins (which are hard coded into DFSYNTHE). Such a threefold speed up is especially important for 1.5D calculations which require  an immense number of 1D calculations (about a million per cube with 1000 times 1000 horizontal grid points).

In this context, we implemented a flexible treatment of the bin grid, enabling the user to choose the overall wavelength interval on which the opacity is calculated (keeping the maximum range between 9 nm and 10 000 nm), as well as the binning and sub-binning on this interval (more details on the implementation can be found in Appendix~\ref{app:example_input}). The underlying high-resolution opacity is then only calculated in the chosen wavelength interval to save computational time. The flexible wavelength binning is in particularly useful if synthesised spectra need to be compared to observations of different resolving power. In order to reduce the computational cost, the number of sub-bins per bin can be reduced significantly if an optimal sub-bin border distribution is chosen. In addition, having a flexible binning and sub-binning grid allows  rapid computation of ODF tables for broadband filters if only the flux through a particular filter is needed \citep{Cernetic_sub, Anusha_2020}.

\begin{figure}
  \centering 
 {\includegraphics[width=0.95\linewidth]{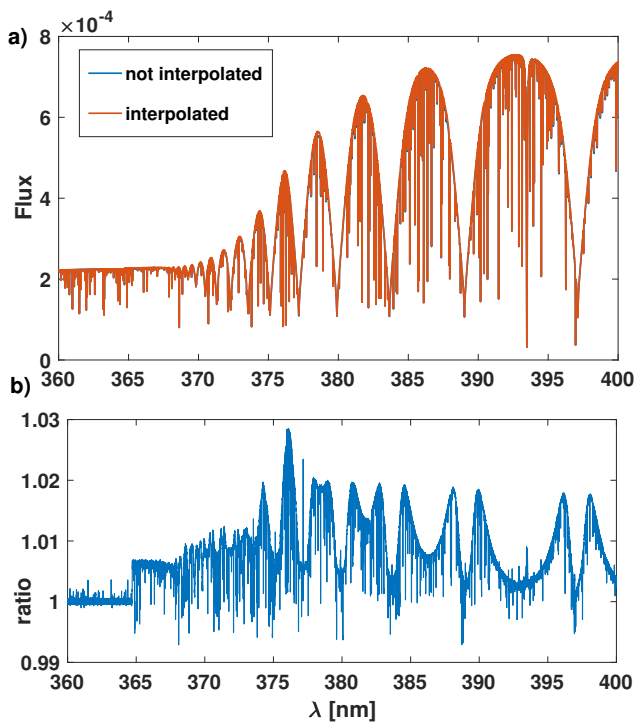}}
  \caption{High-resolution emergent flux for Vega around the Balmer jump region, in the range of 360 nm to 400 nm. a) Flux calculated using interpolated opacities from  pre-tabulated high-resolution opacities on a T-P grid and using opacities calculated for each depth point. b) Ratio of the fluxes from a). }
  \label{fig:interpolated_vs_not}
\end{figure}

\begin{figure}
  \centering 
 {\includegraphics[width=0.95\linewidth]{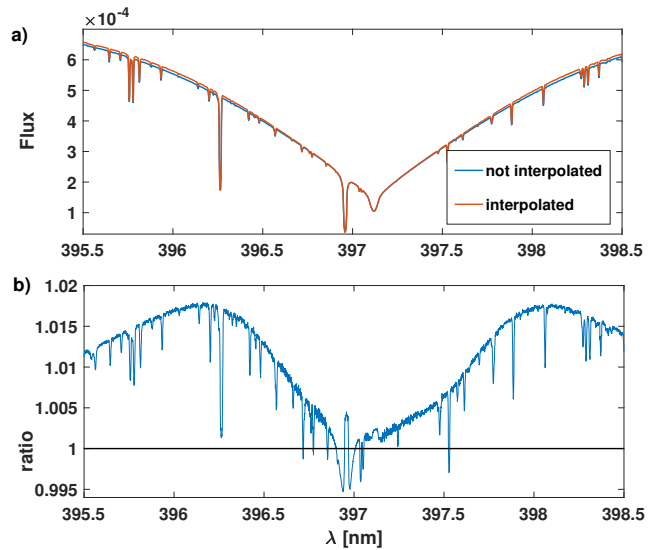}}
  \caption{Same as Fig.~\ref{fig:interpolated_vs_not} for a smaller wavelength interval (395.5 nm to 398.5 nm), corresponding to the $\rm H \varepsilon$ line core.}
  \label{fig:interpolated_vs_not_zoom}
\end{figure}

Moreover, we have added an option to pre-tabulate the opacity on a high-resolution grid.
For that, the code reads the starting and end wavelength of an interval, and uses a resolving power of R=500 000 in this interval. The opacity is then written in the same format as an ODF, but using only one sub-bin, which contains the opacity of the wavelength point. This can be used to calculate emergent intensity at high spectral resolution.
Since the line opacity on a high resolution grid might have strong changes with temperature and pressure, the interpolation error is potentially greater compared to an ODF. An exemplary  calculations of the Vega flux in the wavelength region of the Balmer jump shows that the error between the flux using  the interpolated opacity (using a T-P grid that splits the same ranges as used in  \citet{Castelli_2005_DFSYNTHE} into 100 logarithmically equidistant steps) and the calculated high-resolution opacity  can reach up to 3\% (see Fig.~\ref{fig:interpolated_vs_not}). The largest deviations occur in line wings of strong lines (see Fig.~\ref{fig:interpolated_vs_not_zoom}), which indicates that the  temperature steps  in the considered T-P grid are insufficient to account for the line broadening sensitivity. Therefore, the T-P grid can be adjusted depending on the needed accuracy (for more details see Subsec.~\ref{subsec:flex_tp_grid}). 


\subsection{Flexible T-P grid for pre-tabulation} 
 \label{subsec:flex_tp_grid}
 The DFSYNTHE code  uses a predefined T-P grid of 57 temperature and 25 pressure values for pre-tabulating ODF tables. The pressure covers 12 orders of magnitude ranging from $10^{-4}$ to $10^8$, and the temperature ranges from $2\times 10^{3} -2 \times10^{6}$ K. In MPS-ATLAS the number of T-P points and their values can be specified (as input parameters). This has two advantages. On the one hand, the numerical cost can be reduced, while the resolution of the T-P grid is kept, but the ranges of  temperatures and pressures are adjusted to certain types of stars. On the other hand, the resolution can be increased if more accurate interpolation is needed.  This is especially important if the code is used to generate pre-tabulated high resolution opacity (see Fig.~\ref{fig:interpolated_vs_not}).

\subsection{Line pre-selection criterion} 

\begin{figure}
    \centering
    \includegraphics[width=1.0\linewidth]{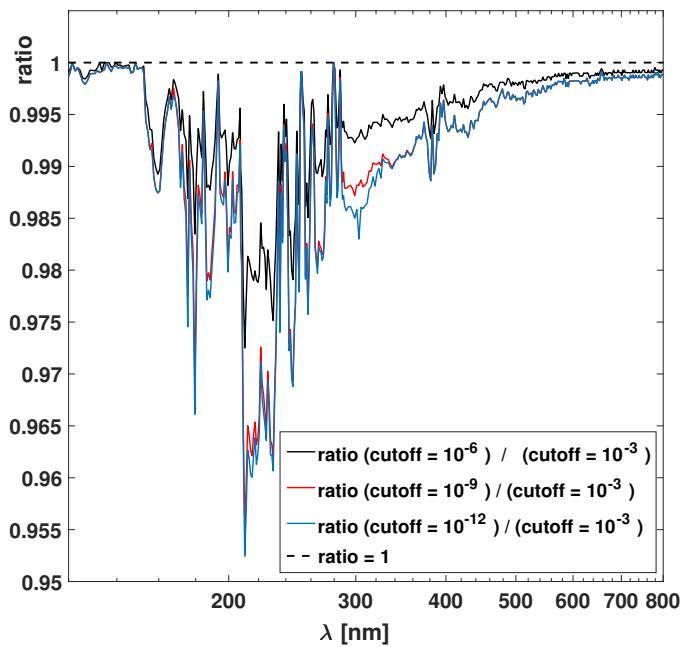}
    \caption{Ratios of the solar emergent intensity at the disc centre calculated using ODF tables that were generated with different cutoff factors for the line pre-selection. Here cutoff factor refers to the opacity of a given line relative to the continuum opacity. If this ratio drops below the cutoff factor, this particular line is not included in the computation of the intensity spectrum.  The default value of the cutoff factor is $10^{-3}$. }
    \label{fig:Limit_ODFline}
\end{figure}

The most extensive line lists such as Kurucz's original line list\footnote{http://wwwuser.oats.inaf.it/castelli/sources/dfsynthe.html} and VALD 3 \citep{VALD_2015} contain many lines that do not contribute to the opacity within the temperature and pressure range of the ODF table. To reduce the computational cost in the line opacity calculations, the lines are  pre-selected based on their line core opacity and the continuous opacity for each T-P point. For that, lines which lead to an opacity in the line core of several orders of magnitude (controlled by the cutoff factor)  less than the continuous opacity are discarded, see Appendix~\ref{app:ODF-impl} for the selection criteria.  
By default the cutoff factor is set to $10^{-3}$. Here, we tested how the solar surface intensity is affected by choosing several smaller cutoff factors (see Fig. \ref{fig:Limit_ODFline}). Since a smaller cutoff factor increases the computational time significantly, the smallest value we tested is $10^{-12}$. The difference between a cutoff factor of $10^{-9}$ and  $10^{-12}$ is negligible. However, overall the largest decrease in intensity can be observed around 200 nm, where it reaches up to 5\%. This implies that even very weak lines have a non-negligible effect on the flux, in particular if there are many of them. 
%

\begin{figure*}
 \sidecaption
  \includegraphics[width=12cm]{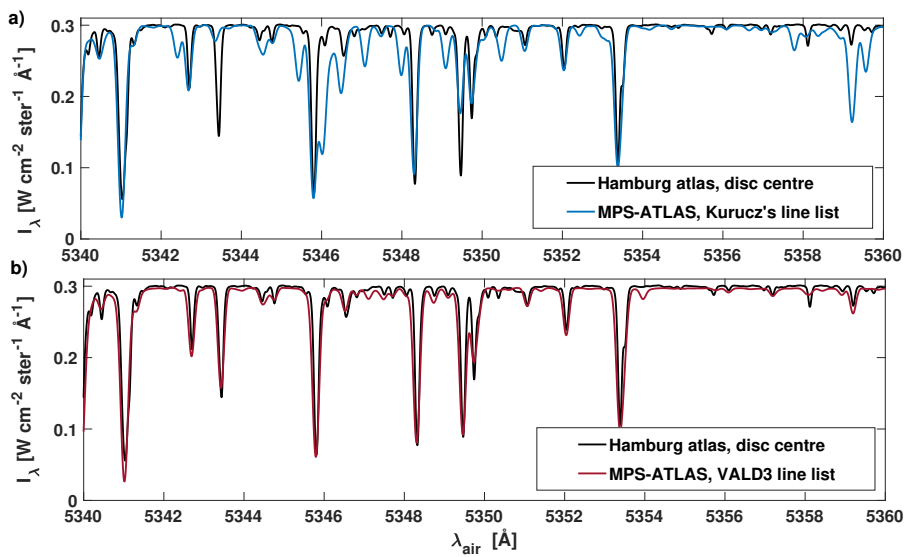}
  \caption{High-resolution solar disc-centre intensity in the range 5340\AA --5360\AA~ computed with MPS-ATLAS using different line lists together with data from the Hamburg atlas of the solar spectrum \citep{Neckel_1999kpatlas, Doerr_2016_KPatlas}:  a) computed intensity using Kurucz's line list, and the Hamburg atlas data, b) intensity using the VALD3 line list, and the Hamburg atlas data.}
  \label{fig:VALD_vs_Kurucz_hr}
\end{figure*}

\begin{figure}
  \centering 
   {\includegraphics[width=1.0\linewidth]{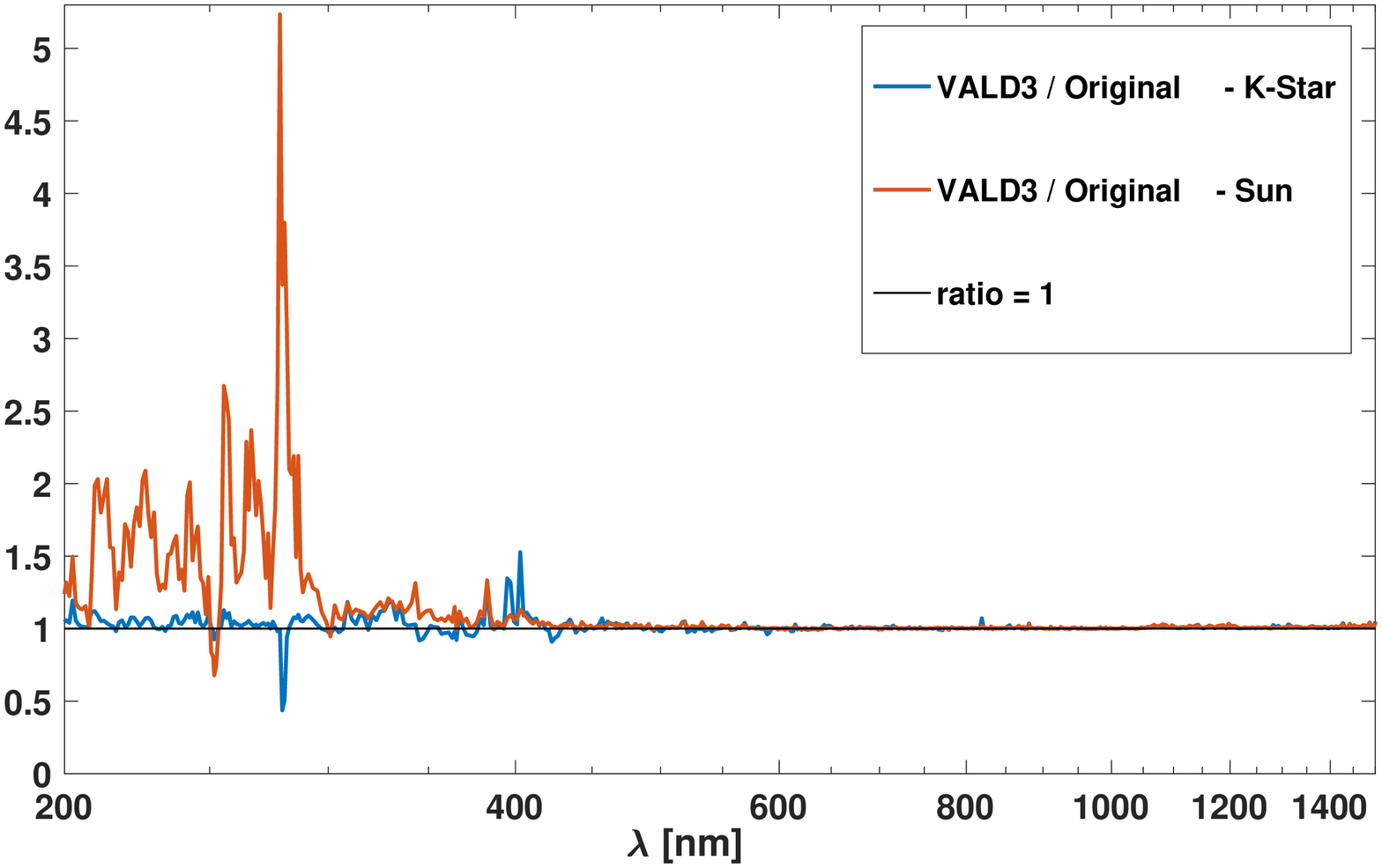}}
  \caption{Ratios of the emergent intensity at disc-centre  calculated using ODF tables with the VALD3 line list to the emergent intensity at disc-centre calculated using ODF tables with  Kurucz's original line list. }
  \label{fig:ratioVALD_vs_Kurucz}
\end{figure}

\subsection{Changing line lists} 
 Kurucz's original line list includes an immense number of theoretically computed lines which have not been measured experimentally. Over the last few decades the atomic and molecular data has been constantly updated to obtain more accurate data. Recent developments led to various line lists which are updated regularly. In particular there have been several important updates of  the Vienna Atomic Line Database (VALD), which is a database focusing on atomic data  relevant for modelling of stellar atmospheres  \citep{VALD_2015}. 

While the DFSYNTHE code only works with Kurucz's line list,  MPS-ATLAS can be used with different line lists. For the current tests we exchanged the atomic line list, but we kept the line list for diatomic and  H2O lines \citep{Sch1998} as in the original DFSYTHE version. 

To illustrate the effect on the flux caused by the differences in the line lists, 
we first calculated the model atmosphere for the Sun  using the standard `little' grid ODF \citep{Castelli_2005_DFSYNTHE} with the original line list and the VALD3 line list. Then, using these models,  we calculated the intensity at disc-centre on a high-resolution wavelength grid with resolving power R=500 000 for each of the two line lists.  Figure~\ref{fig:VALD_vs_Kurucz_hr} shows both intensities in the range between {5340 \AA --5360 \AA} together with the Hamburg atlases of the solar spectrum \footnote{{ftp://ftp.hs.uni-hamburg.de/pub/outgoing/FTS-Atlas/}} \citep{Neckel_1999kpatlas, Doerr_2016_KPatlas}. For this comparison, we  degraded the resolving power of the calculated flux using a Gaussian kernel. 
It becomes evident that the intensities obtained using the VALD3 line list agree significantly better with the solar observations, while the intensities obtained using  Kurucz's original line list show quite a few lines that are not present in the solar spectrum. This is because Kurucz's line list contains significantly more lines than VALD3 and most of these excess lines have never been measured in the laboratory. 


In a second step, we compare the overall energy distribution for the Sun, and a K-type star ($\rm T_eff = 4000 K $) obtained using different line lists.
For that, we calculated the model atmospheres using  the `little' grid ODF  with different line lists, and subsequently generated the overall disc integrated emergent flux using  the standard `little' grid ODF with the corresponding line list.
Figure~\ref{fig:ratioVALD_vs_Kurucz} shows that the fluxes calculated with the original Kurucz's line list are significantly different from the fluxes obtained using the VALD3 line lists below 450 nm. The effect in the UV is huge for the Sun, where the flux obtained using the original Kurucz's line list gives a better agreement (see a more detailed discussion in Sect.~\ref{subsec:code-to-observ}). However,  for a K type star for which line opacity is even more important than for the Sun there is rather a moderate difference. This indicates that the main difference between the VALD3 and the  Kurucz's line lists are lines whose lower level have higher energies (i.e. mainly lines contributing to opacity at higher temperatures than in K-stars).
 

\subsection{NH photodissociation} 
Comprehensive comparison of observed and modelled solar spectra showed that molecular photodissociation is an important source of continuum opacity in the UV range \citep{Fontenla_2011}. Moreover, \citet{Fontenla_2015} proposed that NH photodissociation is the source of the missing opacity in the UV \citep[see, e.g., ][for the detailed discussion of the missing opacity problem]{busaetal2001, shorthauschildt2009, shapiroetal2010}. So far only the molecules CH and OH were included in the ATLAS9 continuous opacity calculations \citep{Castelli_ATLAS12_2005}.  We added the opacity contribution from the NH molecule using cross-sections that we obtained from  \citet{Priv_Comm_Kurucz}. A more detailed description on how molecules are included in MPS-ATLAS is given in Appendix~\ref{app:solve_Enumbers}, and how CH, OH, and NH photo-dissociation opacities are implemented in Appendix~\ref{app:molec_opacitiy}. This additional opacity source changes the specific emergent intensity by at most $0.5\%$  (see Fig.~\ref{fig:w-no-NH}). We conclude that NH is not sufficient to explain the missing opacity in the UV. 


\subsection{MPI parallelisation for faster computations} 
While the RT calculations can be sped up by using optimised ODF configurations, the calculations of the ODFs table itself remain time-consuming. To achieve faster calculations we parallelised the module for ODF generation (\mbox{module I} in Fig.~\ref{fig:0101}) along temperature values. 
We note that the number of lines  contributing noticeably to opacity depends on temperature. In particular, more lines have to be calculated for lower temperature values so that producing ODFs for them is the most time consuming.
To account for this imbalance of computational time, we used a master-slave implementation. 
Here, one core distributes the T values for which the ODF must be calculated to all other cores as soon as they finish with their previous values. Consequently, while different cores calculate ODFs for different numbers of temperature values, the number of calculations is distributed between the cores more or less equally.
This also has the advantage that the number of T values does not need to be  divisible by the number of cores used. With this implementation, an example ODF table on the same T-P grid as used in the original DFSYNTHE code can be calculated in under 10 minutes on 36 cores on a modern high performance cluster.

%

For the RE calculations (\mbox{module II} in Fig.~\ref{fig:0101}), the integrated flux over the whole spectrum is needed, where the number of frequency points can be varied.
Calculations with a large number of frequency points, or if the  high-resolution grid is used, are very time consuming as the RT needs to be solved in each iteration on chosen frequency grid.   Thus,  we also parallelised the RT calculations along  wavelengths in the module for modelling atmosphere structures.

Finally, module III that calculates the emergent intensity (see \mbox{module III} in Fig.~\ref{fig:0101}) uses a parallelisation if the intensity for more than one model has to be calculated. This is only the case for post-processing 3D calculations. In this case the atmosphere models are distributed among the available cores.

\subsection{Speed-up \& portability} 
\begin{figure}
    \centering
 \includegraphics[width=1.0\linewidth]{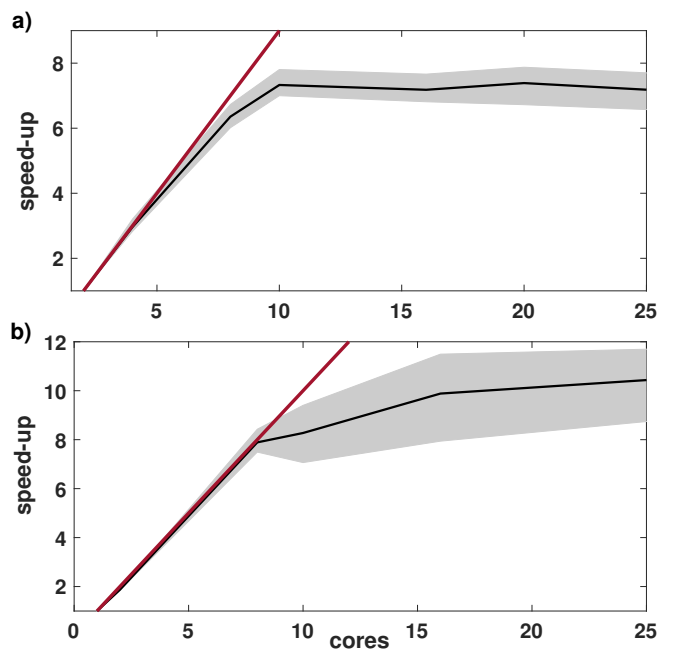}
    \caption{Top panel: Speed-up for ODF table generation as a function of the number of compute cores used (module I from Fig.~\ref{fig:0101}). Bottom panel: Speed-up for atmosphere model calculation (module II from Fig.~\ref{fig:0101}).  Black lines in both panels indicates the speed-up averaged over 5 runs.  The grey shaded area indicates the spread in the 5 runs, and the red line indicates the ideal speed-up.}
    \label{fig:scale_ODF_model}
\end{figure}
Using the parallelised version of the ODF module we tested the MPS-ATLAS scalability. For the ODF calculation we used the standard T-P grid as in \citet{Castelli_2005_DFSYNTHE}, but set the wavelength interval from 200 nm -- 10 000 nm, which significantly reduces the computational time. The low temperature values in the standard grid (below 5000 K) require the highest number of lines, and thus are very-time consuming compared to larger temperature values.

Fig.~\ref{fig:scale_ODF_model} shows the speed-up for ODF generation and RE calculations. It becomes evident that there is no further speed-up of the ODF calculations for more than 10 cores. This is because the lowest temperature value (1995~K) forms the bottleneck in this example. Even if all other temperature values are calculated faster using more cores, at the end of the entire ODF calculation 
the temperature value that takes the longest determines the amount of time needed. The number of cores at which such a bottleneck (due to one particular temperature)   appears depends on the T-P grid, and wavelength range of the ODF table. In general, the fewer P points and more T points a grid has, the higher the number of cores until the speed-up saturates.

To test the parallelisation for the RE calculations, we chose to model a 4000 K star with $\rm log g = 4.8$, and M/H = 0.3. We used a standard `little' grid ODF \citep{Castelli_2005_DFSYNTHE},  and reduced the wavelength interval to the first 1210 points on the grid, which corresponds to 9 nm -- 10 000 nm. The model calculation converges after 37 iterations.   
The speed-up of the  RE calculations (\mbox{module II}) shows a different behaviour: the increase in speed-up starts to flatten after 8 cores (see Fig.~\ref{fig:scale_ODF_model} b). The flattening happens because only the part of the RT calculations along the wavelength points is parallelised, while all other calculations, especially the temperature corrections are executed on one core. Here, the behaviour of the speed-up depends on the number of wavelength points that need to be calculated.  

The parallelisation for the emergent intensity calculations (\mbox{module III}) is along atmosphere models, thus it is only useful for calculating along model atmospheres from 3D cubes.  Since it only needs a few communications in the beginning to   distribute the input settings, we expect an almost ideal speed up. 

The code is fully portable. The available git version of the code includes an installation script that downloads all necessary libraries and files, such as a NetCDF library, line lists, and example input files. The code can be compiled using an openMPI and intel compiler, as well as a corresponding gnu compiler. Note that for the gnu version, a different line list format is needed, which we provide on request.

\section{Benchmark} 
\label{Sec:Code_testing}
In order to put the MPS-ATLAS code into context, we compared emergent flux obtained by the MPS-ATLAS code to that from other codes and to observations. 
For the code-to-code comparison the stellar parameters were chosen to be available on both the PHOENIX- and Kurucz- model grid of calculated specific intensities. To this end, we considered  three  hypothetical stars of spectral classes A, F, and K (with corresponding temperatures of 8000 K, 6500 K, and 4000 K). These spectral classes have been chosen to test the performance of the code for cases when opacity is dominated by different sources. 
For example, in a K-type star the opacity is mainly  due to by millions of atomic and molecular lines while  the main opacity source in an A-type star  is  the continuum and hydrogen lines. Furthermore, for comparison to observations, we used the Sun and Vega (see Table~\ref{table:01} for the summary of the fundamental stellar parameters of the stars we used). We chose Vega for comparison, because accurate measurements of the total flux over a large wavelength range are available.

\begin{table}
\setlength\tabcolsep{5pt}
\renewcommand{\arraystretch}{1.5}
\caption{Stellar parameters for the comparison stars.}              
\label{table:01}      
\centering                                     
\begin{tabular}{|c | c| c| c| c|c| }         
\hline\hline                       
star        & A-type & F-type & K-type & Sun & Vega  \\   
\hline\hline                                 
 $\rm T_{eff}$ in K & 8000  &  6500 & 4000 & 5777 & 9550  \\
 \hline\hline                                 
 $\rm log$ g & 4.0  &  4.5 & 5.0 & 4.335 & 3.90  \\
  \hline\hline                                 
 M/H & 0.0  &  0.0 & 0.0 & 0.0  & -0.5 \\
\hline\hline                                             
\end{tabular}
\end{table}

\subsection{Code-to-code comparison}
\label{subsec:code_to_code}
We compare emergent fluxes calculated using the MPS-ATLAS code to the original fluxes from the ATLAS9 grid calculated by Kurucz\footnote{http://kurucz.harvard.edu/grids.html}  \citep{1993KurCD..13.....K},  and to the  modelled fluxes on the PHOENIX grid \footnote{http://phoenix.astro.physik.uni-goettingen.de} for three different stellar types (K-type, F-type and A-type; see Tabel~\ref{table:01} for details). In our calculations we assume the same setting as  in the Kurucz grid modelled fluxes:  convection is turned on, without overshoot, and the mixing length is set to 1.25. We further consider a constant micro-turbulence of \mbox{2 km/s}. Generally, the PHOENIX code provides a more intricate setup, e.g. the models are calculated in spherical geometry, condensation is included in the equation of state, and some NLTE effects can be turned on.  
However, the model grid we use \citep{Phoenix_descript2013} was obtained using LTE and the RT calculations only make use of the NLTE effects by a special line profile treatment for some species (Li I, Na I, K I, Ca I, Ca II). Moreover, the mixing length  and the micro-turbulence parameters vary on the PHOENIX model grid, and thus can slightly differ from the one used in ATLAS9 and MPS-ATLAS \citep[for more details see][]{Phoenix_descript2013}.   

We preformed calculations using two different elemental compositions.  For a comparable calculation to the Kurucz grid we used the same element abundances as in Kurucz's calculations. These abundances are taken from \citet{Anders_Grevesse_1989}, and will be referred to as `Anders composition'. For the comparison with PHOENIX we used the more up-to-date `Asplund composition' taken from \citet{Asplund_2009}.  
For the model-to-model comparisons, both the model atmosphere and the emergent spectra are calculated using the standard `little' ODF with 1221 frequency points \citep{Castelli_2005_DFSYNTHE}. 
We smoothed all spectral fluxes by applying the average over a 15 nm interval around each wavelength grid point. The PHOENIX high-resolution spectral fluxes are first averaged over the same intervals as given by the ATLAS wavelength grid, and then smoothed with the same procedure over 15 nm intervals.

%
%
\begin{figure}
  \centering 
  {\includegraphics[width=1.0\linewidth]{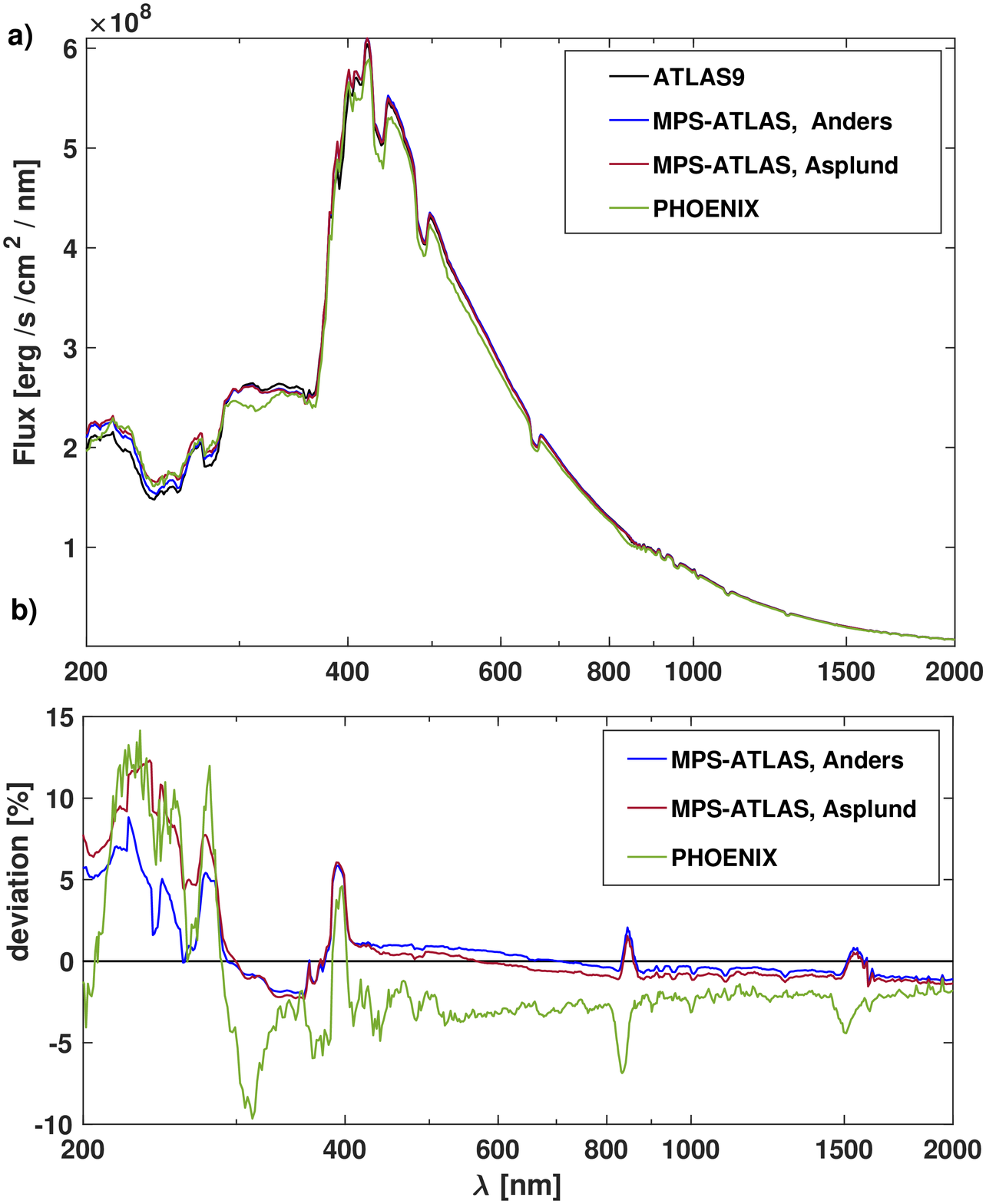}
  }
  \caption{Flux comparison between MPS-ATLAS, Kurucz-ATLAS9 and PHOENIX code for an A-type star with the effective temperature of 8000K.   Flux values are shown (top panel) together with the corresponding flux deviations (bottom panel) in \% compared to the original Kurucz calculations. }
  \label{fig:A-type-star}
\end{figure}

\begin{figure}
  \centering 
  {\includegraphics[width=1.0\linewidth]{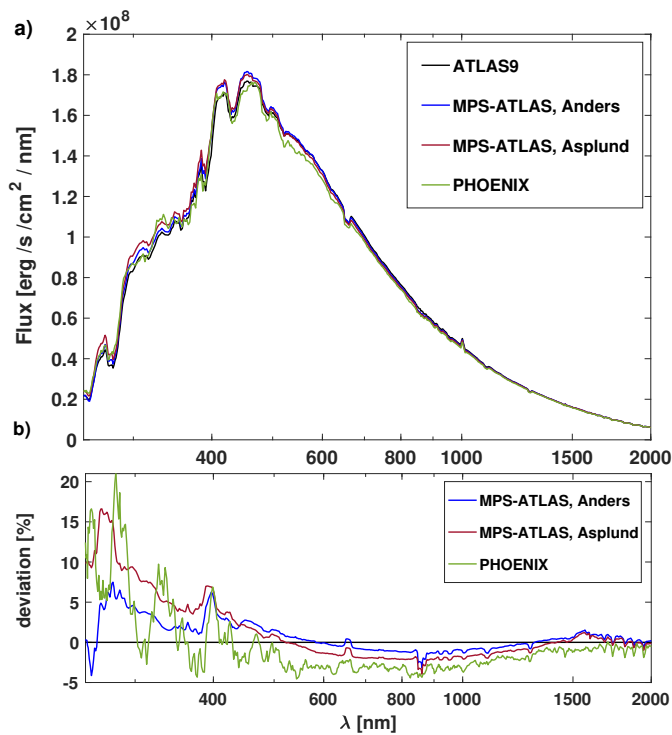}
  }
  \caption{Same as in Fig.~\ref{fig:A-type-star}, but for a F-type star with $\rm T_{eff}= 6500$K. }
  \label{fig:F-type-star}
\end{figure}

\begin{figure}
  \centering 
  {\includegraphics[width=1.0\linewidth]{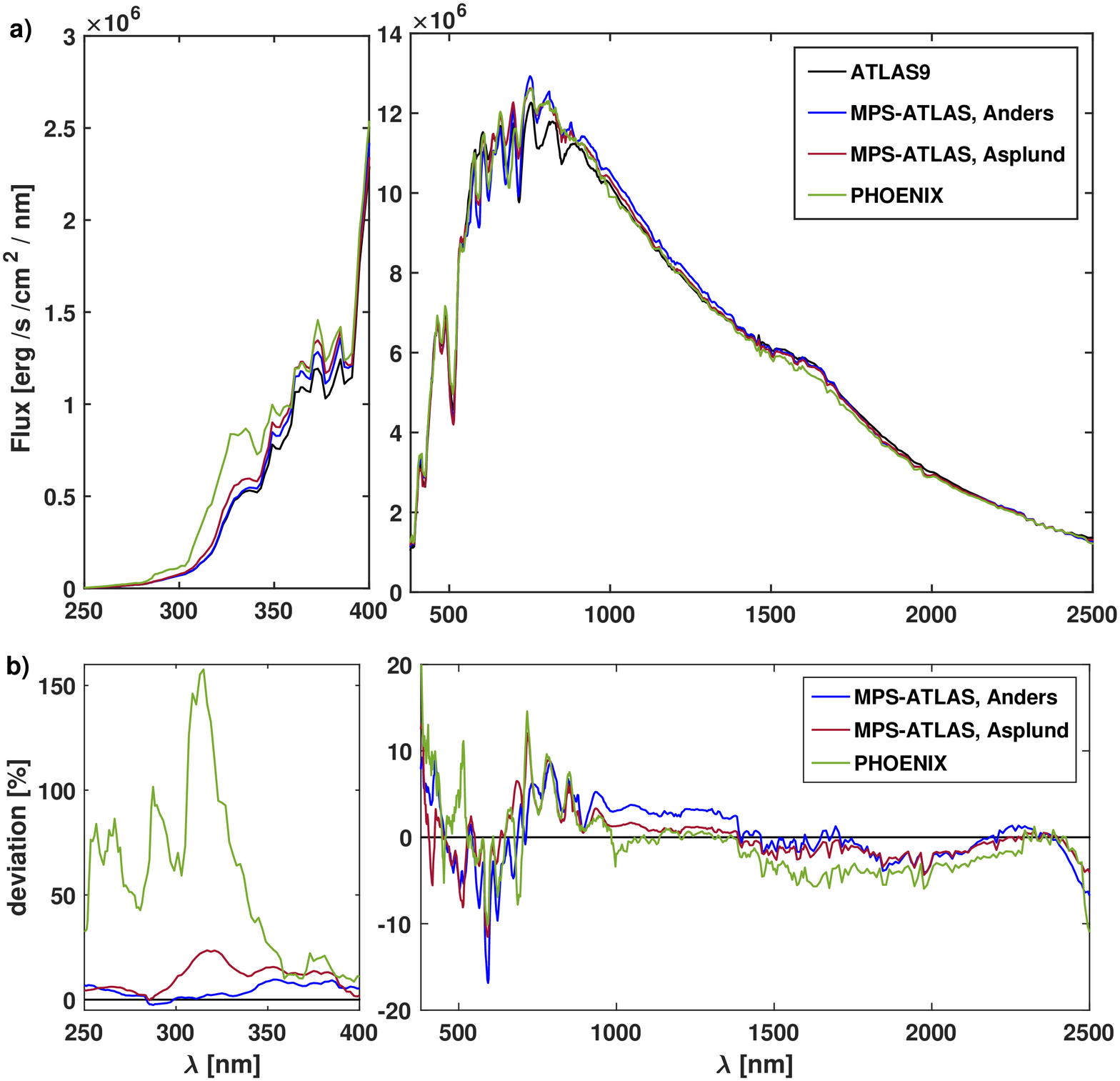}
  }
  \caption{Flux comparison between MPS-ATLAS, Kurucz-ATLAS9 and PHOENIX code for a K-type star with $\rm T_{eff}= 4000$K. Flux values are shown (the two top panels) together with the corresponding flux deviations (in the two bottom panels) in \% compared to the original Kurucz calculations.}
  \label{fig:K-type-star}
\end{figure}

For an A-type star the emergent spectral flux returned by the MPS-ATLAS code and those from PHOENIX and ATLAS9 grids are shown in Fig.~\ref{fig:A-type-star}a. For a more detailed  comparison we show the deviations of the calculated MPS-ATLAS fluxes, and the PHOENIX to the original ATLAS9 fluxes in Fig~\ref{fig:A-type-star}b. 
Overall there is a reasonable agreement between the three codes. The largest deviations are in the UV for wavelength shorter than 400 nm. A good example for the importance of the elemental composition can be seen in the range between 210 nm and 290 nm. While the flux obtained by using the MPS-ATLAS code and the `Anders composition' is closer to the ATLAS9 calculations (also performed with the `Anders composition'), the MPS-ATLAS flux for the `Asplund composition' is closer to the PHOENIX calculations (performed with the `Asplund composition'). Large differences can occur especially in the UV due to different line lists used.  In the wavelength interval between 300 nm - 1000 nm the PHOENIX calculation gives an overall smaller spectral flux than fluxes calculated with MPS-ATLAS and ATLAS9. The difference is approximately  5\%  around 400 nm and then decreases to a few percent for wavelengths greater than 1000 nm. In this wavelength interval the dominant continuum opacity contributor are $\rm H^- $ bound-free transitions. Thus either a slight difference in the equilibrium number densities or a different implementation of the  $\rm H^- $ bound-free cross-section can potentially lead to these differences. 

\begin{figure*}
\sidecaption
  \includegraphics[width=12.5cm]{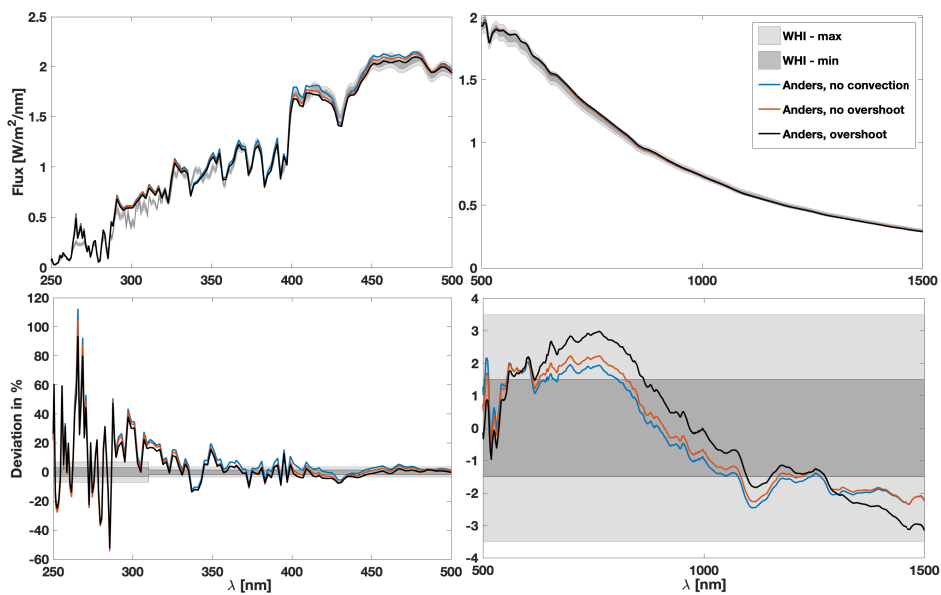}
    \caption{Irradiance calculated for solar parameters compared to WHI.
    The light and dark grey shaded area indicates the minimum and maximum measurement error of the SIRS WHI (Solar Irradiance Reference Spectra for the 2008 Whole
Heliosphere Interval), respectively \citep[see,][for a detailed description]{Woods_2009_WHI}.  Solar irradiance (top panels) calculated using MPS-ATLAS with three different assumptions: no convection, with convection but no overshoot, and with convection and overshoot.  The flux deviations between WHI observed irradiance and models in \% are shown in the bottom panels.}
    \label{fig:whi-first-conv}
\end{figure*}

In addition there are three larger deviations around 820 nm, 1458nm, and 2279nm which correspond to the Paschen limit, the Brackett limit and the Pfund limit, respectively. This implies that all three codes have a different treatment of the hydrogen continuous transitions of these series.  On the contrary, the Balmer series transitions, H-$\alpha$ (at 656nm)  to H-$\zeta$ (at 389nm), show only very slight differences  between the three codes. The significant deviations around 395nm are caused by the Ca II H and K doublet, where the MPS-ATLAS calculation gives more similar results to  PHOENIX than to the older ATLAS9 calculations. We note, however, that the proper treatment of the Ca II H and K doublet  requires  NLTE modelling which is absent in all three codes.

In Fig.~\ref{fig:F-type-star} the emergent flux for a F-type star (see  Table~\ref{table:01} for exact fundamental parameters) is displayed. Overall the deviations in the UV are greater and reach up to 20\%, while the differences in the visible and infrared are smaller. 
The two elemental compositions entering the MPS-ATLAS lead to the following differences in the spectra in the range 200 nm -- 500 nm: the flux is larger for the `Asplund composition', and it decreases towards 450 nm whereafter the flux is smaller for the `Asplund composition' than for the `Anders composition'.  This behaviour is most probably caused, directly or indirectly, by the effect of the composition on line blanketing. The line opacity, in particular from iron,  is smaller for the 
`Asplund composition' and allows more photons to escape. Then, the decrease in the visible can be explained by a compensation effect as the wavelength integrated flux has to be the same for both element compositions. 
The PHOENIX flux oscillates around the MPS-ATLAS flux obtained using the `Anders composition' in the region 200 nm -- 400 nm, which implies that the line opacity  differs significantly due to a different line list.  
Finally, the wave-like deviation between 500 nm --  2600 nm indicates that the photospheric structures differ somewhat from each other. It becomes evident that the PHOENIX, and MPS-ATLAS model have a similar structure, while the older Kurucz's model deviates. 
%

The deviations between the codes are larger for a K-type star (see Fig.~\ref{fig:K-type-star}). In the UV the difference between PHOENIX and other codes reaches 160\%, this might be due to a smaller number of UV lines contributing to opacity at lower temperatures (i.e. below 4000 K) in the line list used in the PHOENIX code. The flux calculated using  MPS-ATLAS deviates by  up to 25\% from Kurucz's originally computed flux.  Focusing on wavelengths longward  of 400 nm, the deviations oscillate, but the amplitude of the deviations decreases towards the infrared.  Comparing the difference between the `Anders composition' and the `Asplund composition', the flux obtained using the `Asplund composition' is closer to the PHOENIX calculation for most of the wavelength in the visible. This is also the case in the infrared, where the overall deviations become less than 10\% between all four models.


\subsection{Code-to-observation comparison}
\label{subsec:code-to-observ}
\begin{figure*}
\sidecaption
  \includegraphics[width=12.5cm]{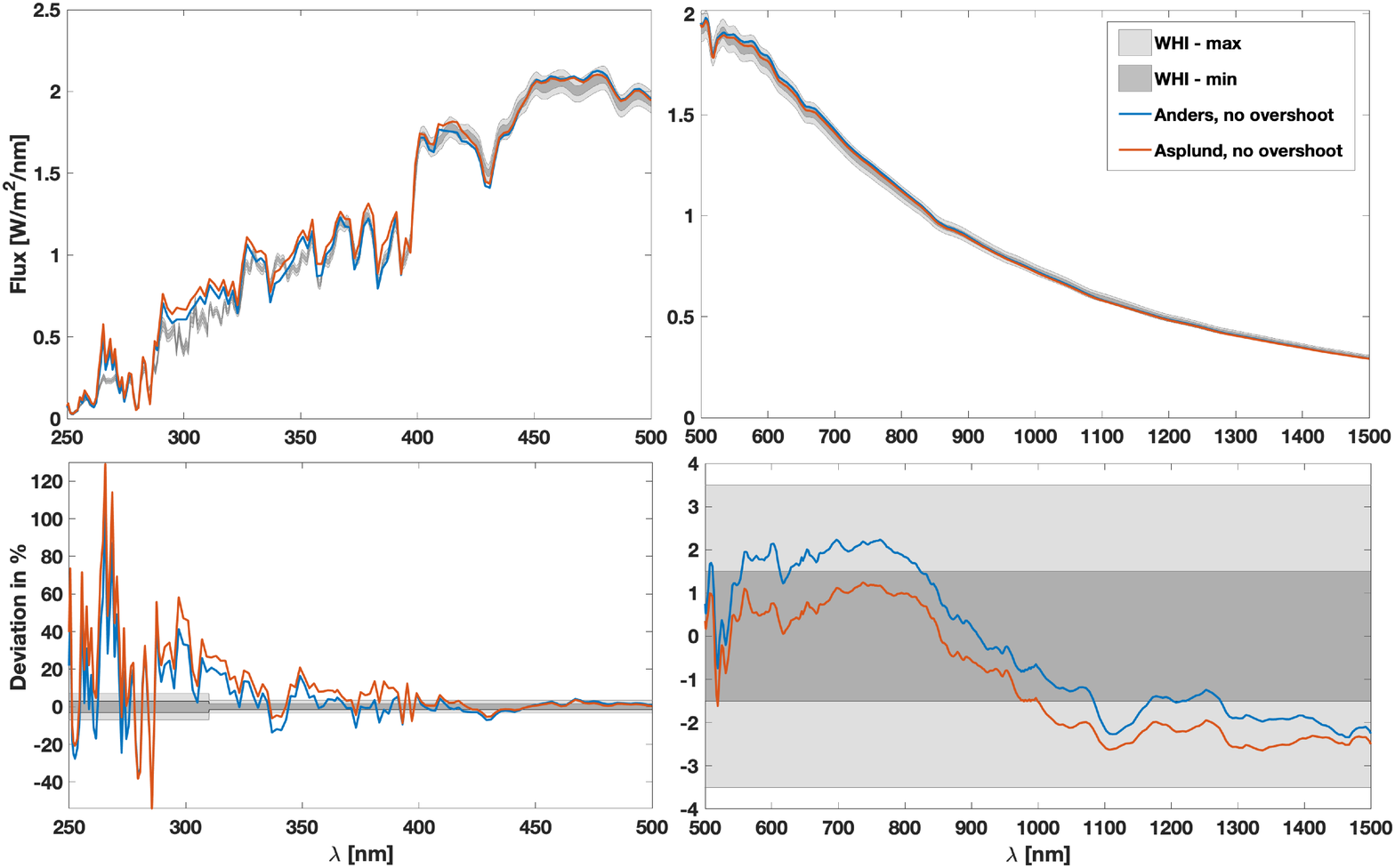}
    \caption{Irradiance calculated for solar parameters compared to WHI. Same as in Fig.~\ref{fig:whi-first-conv}, except that the MPS-ATLAS model calculations were done without overshoot and using the Anders and Asplund elemental compositions. }
    \label{fig:comp_model-whi_element}
\end{figure*}

\begin{figure*}
\sidecaption
  \includegraphics[width=12.5cm]{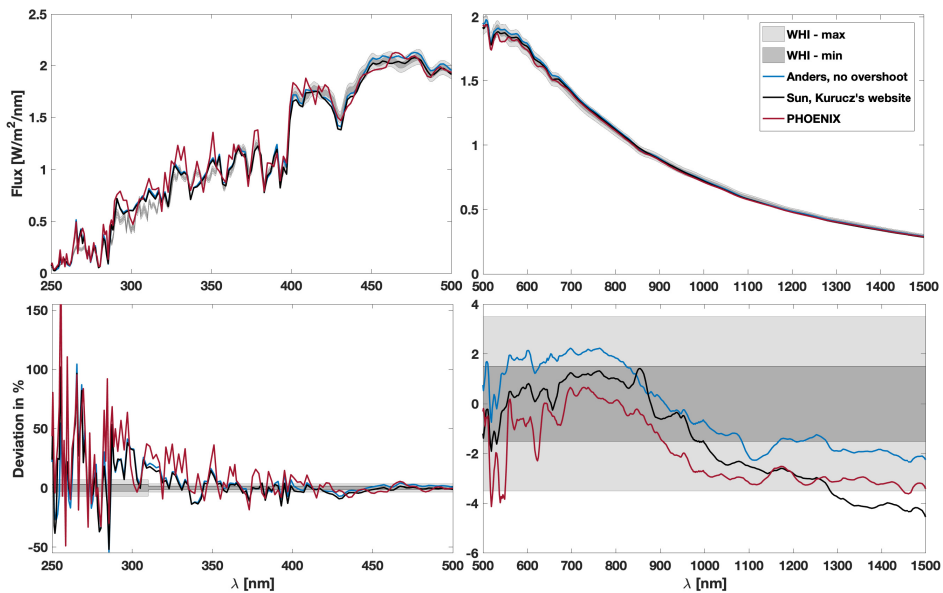}
    \caption{Irradiance calculated for solar parameters compared to WHI. Same as in Fig.~\ref{fig:whi-first-conv}  , but without the Asplund composition, while in addition the solar flux provided on Kurucz's website and the flux from PHOENIX  are shown.}
    \label{fig:comp_model-whi}
\end{figure*}

Besides comparing with the output of other codes, it is important to also check how well MPS-ATLAS reproduces observations.  For the Sun a lot of accurate high-resolution data exist, as well as low resolution spectra. In contrast, for most of the stars either broadband fluxes, intermediate resolution spectra or normalised (e.g. to continuum level) high-resolution spectra for a rather small wavelength interval are available.  Here we first test how well the solar flux can be matched by the MPS-ATLAS code, and in a second step we chose Vega as a comparison star.

\subsubsection{The Sun}
%
For the calculations presented in this subsection both the model atmosphere and the emergent spectra are calculated using the  standard `little' ODF with 1221 frequency bins \citep{Castelli_2005_DFSYNTHE}. For all considered cases  we recalculated the model when changing any parameters. We always assumed a micro-turbulence of 1.5 km/s and if convection was turned on, the mixing-length is 1.25. 
We compare our calculations to Solar Irradiance Reference Spectra (SIRS) for the 2008 Whole Heliosphere Interval \citep[WHI][]{Woods_2009_WHI}. For a better comparison all calculated fluxes are smoothed out using a trapezoidal kernel (Harder, private comm.) to match the WHI resolution in wavelength regions where the `little' ODF resolution is higher than that of the WHI. We kept the original resolution otherwise.

Fig.~\ref{fig:whi-first-conv} shows the solar irradiance at a given wavelength at one AU from the Sun as in the  WHI  compared  to MPS-ATLAS calculations using the `Anders composition'  with and without convection, and overshoot. One can see that in the mid UV the deviations between the calculated and observed irradiance significantly exceed the uncertainty of the measurements.  This is not surprising (and  similar deviations are also present in the ATLAS9 and PHOENIX calculations, see below) since it is well known that currently available line lists miss a significant number of weak lines \citep[this excess flux constitutes the famous `missing UV opacity' problem, see e.g. the recent review by][and citations therein]{Rutten_rev}. We note that the solar UV spectrum is also affected by deviations from LTE, which are ignored in MPS-ATLAS calculations. However, it has been shown that accounting for non-LTE effects will make the deviations between the calculated and observed UV flux level even larger \citep[see, e.g.,][]{shorthauschildt2009, shapiroetal2010, Rinat_2019}.

In contrast to the UV spectral domain, the modelled irradiance agrees very well with the observations in the visible and infrared.
Fig.~\ref{fig:whi-first-conv} shows that the deviations there are mostly within the minimum measurement uncertainty. While the differences between the three calculations are very small, the agreement between the observations and the MPS-ATLAS calculation including convection with overshoot is best in the interval 450 nm -- 650 nm. However, in the region  650 nm -- 900 nm the fluxes including overshoot show the greatest deviations.  The wave-like behaviour of the deviations in the range between 500 nm -- 1500 nm might indicate that the modelled atmosphere temperature  in the region where the continuum is formed does not accurately  match  the actual  temperature in the Sun.

The solar elemental composition has been  intensively investigated for the last decades and updated following more accurate modelling \citep{Anders_Grevesse_1989, Grevesse_Sauval_1998, Asplund_2009}. Changing ratios between hydrogen, helium and heavier elements, which significantly contribute to the electron number density, affect  the structure of the atmosphere and the spectral synthesis in a competing way. Increasing the concentration of the electron donors without re-calculating the atmospheric structure leads to a drop of the flux, and thus a decrease of the effective temperature. However, as soon as the structure is re-calculated in RE, it compensates for the decreased flux and the effective temperature should return to its original value.

In the next step we show the difference in the irradiance for the `Anders composition' and `Asplund composition', both with convection but no overshoot. Fig.~\ref{fig:comp_model-whi_element} shows flux model calculations for the different elemental compositions, together with the WHI measurements. 
The difference in the elemental composition leads to a redistribution of the flux in different wavelength intervals due to a change of equilibrium number densities of the species (which in turn affects the opacity).  The effective temperatures obtained from the total flux are almost identical for the two compositions, being  $\rm T_{eff} = 5778.9 K$ for the `Asplund composition', and  $\rm T_{eff}= 5778.5$ for the 
`Anders composition'. Nevertheless,  the irradiance calculated using the `Asplund composition' has a higher flux in the UV, compared to both the WHI and the flux using the `Anders composition', so that the former displays greater deviations  from the observations. In contrast, the flux calculated using the `Asplund composition' is lower in the visible and IR compared to that calculated using the `Anders composition'. Note, that the same behaviour was also observed for the A-type and F-type star of solar metallicity (see discussion in Subsection~\ref{subsec:code_to_code} and Figs.
~\ref{fig:A-type-star}-\ref{fig:F-type-star}).  Both calculations match the observations within the measurement  uncertainties for wavelengths greater than 450 nm.

Finally, in Fig.~\ref{fig:comp_model-whi} we plot the flux obtained by MPS-ATLAS for the `Anders composition' without overshoot, together with the original fluxes calculated by Kurucz\footnote{http://kurucz.harvard.edu/stars.html}, and the PHOENIX flux calculations. Since the PHOENIX grid does not have the flux for the solar effective temperature, and surface gravity, we used linear interpolation between the closest available stellar parameters. Furthermore, before applying the same smoothing procedure using a trapezoidal kernel to match the WHI resolution, we averaged the PHOENIX flux over the same wavelength intervals as given in the standard `little' ODF grid.   
The interpolated PHOENIX flux shows a slightly worse agreement with the observations in the UV, but an overall good agreement in the visible and IR.  Kurucz's original calculation agrees better with the observations  in the interval 450 nm -- 650 nm than the PHOENIX flux. All three codes show a wave-like behaviour  between 500 nm -- 1500 nm indicating a slight  mismatch of the modelled atmosphere as discussed above. 

Note that we used  Kurucz's original line list for the above atmosphere model and flux calculations. The modelled flux in the UV is significantly  larger than the observed values, and using the VALD3 line list leads to an even greater  flux in the UV (see Fig.~\ref{fig:ratioVALD_vs_Kurucz}). This indicates  that  Kurucz's line list leads to a better agreement of the calculated UV  flux to observations. This is because Kurucz's line list contains significantly more lines than VALD3 and even though most of these lines have never been measured in the laboratory they allow a better representation of the UV line haze than the VALD3 lines.

\subsubsection{Vega}


\begin{figure}
    \centering
    \includegraphics[width=1.0\linewidth]{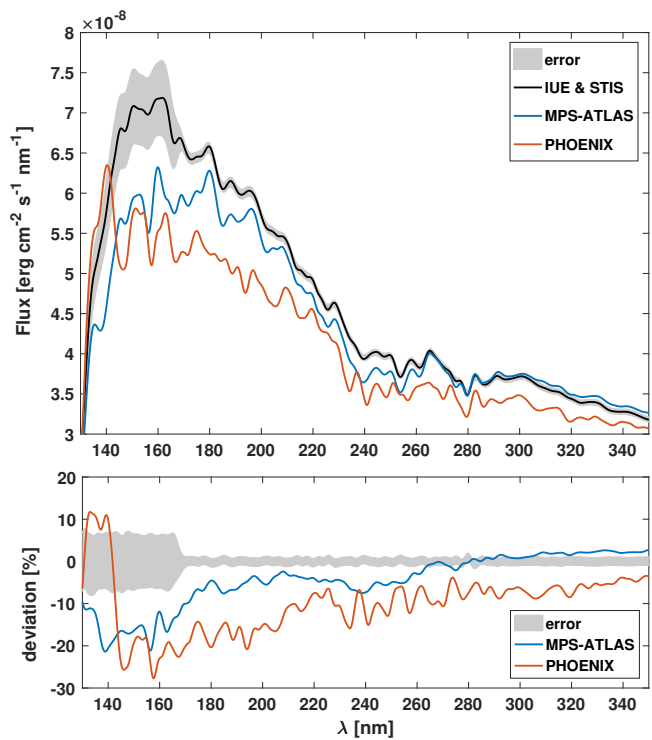}
    \caption{Comparison of measured and calculated UV flux for Vega.
    The grey area indicates systematic and statistical measurement uncertainties. For comparison, MPS-ATLAS flux and PHOENIX flux are plotted. The modelled fluxes are scaled with the same factor, and were obtained for the same effective temperature, metallicity, and surface gravity (Table~\ref{table:01}). Top panel: Absolute flux, with error. Bottom panel:  deviations of modelled fluxes from measured flux in \%.}
    \label{fig:comp_model-Vega-STIS}
\end{figure}

\begin{figure}
    \centering
    \includegraphics[width=1.0\linewidth]{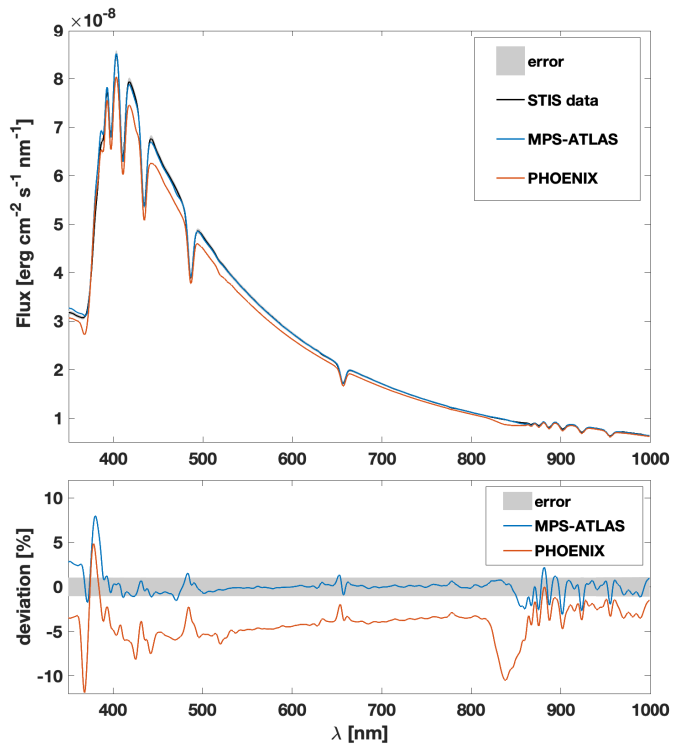}
    \caption{Comparison of measured and calculated flux for Vega. The observed flux consists of data from STIS.
    The grey area indicates systematic and statistical measurement uncertainties. For comparison, MPS-ATLAS flux and PHOENIX flux is plotted. The modelled fluxes are scaled with the same factor, and were obtained for the same effective temperature, metallicity, and surface gravity. Top panel: Absolute flux, with error. Bottom panel:  deviations of modelled fluxes from measured flux in \%.}
    \label{fig:comp_model-Vega-STIS_part2}
\end{figure}

\begin{figure*}
\sidecaption
  \includegraphics[width=12cm]{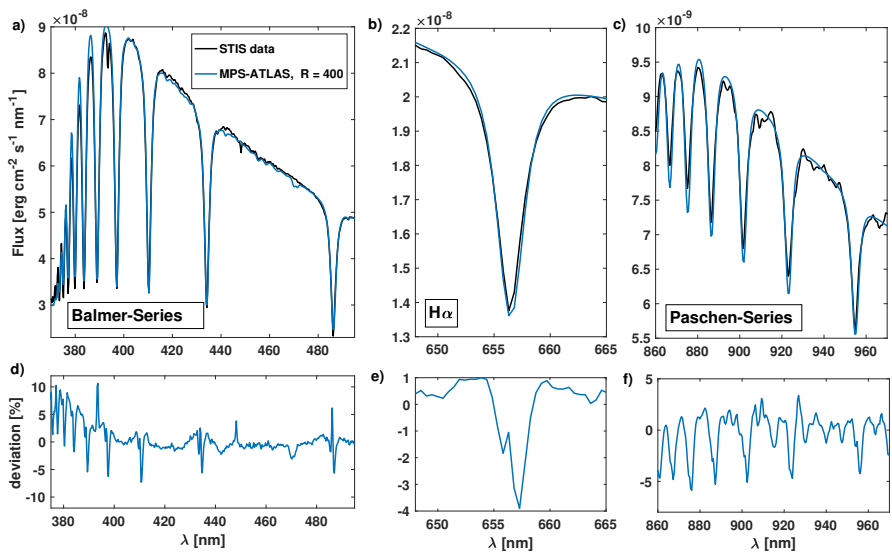}
    \caption{Comparison of STIS flux with MPS-ATLAS calculation with $\rm R = 400$. a) Hydrogen Balmer series. b) $\rm H_{\alpha}$ line. c) Paschen series. d)-f) the deviation of the calculated flux to the measured flux in \%.   }
    \label{fig:Vega_Hlines}
\end{figure*}

The other star for which accurate absolute spectrophotometry is available is Vega. The most up-to-date observed absolute flux consists of data from two instruments, the International Ultraviolet Explorer (IUE), and the Space Telescope Imaging Spectrograph (STIS) mounted on the Hubble Space Telescope \citep{Bohlin_Vega_2004, Bohlin_et_al_2014}. 
The fundamental parameters for Vega still contain a non negligible uncertainty  \citep{1994A&A...281..817C}. Therefore, we tested a small range of stellar parameters to seek good agreement, where the  range  was chosen based on previous stellar parameter determinations \citep{1994A&A...281..817C, 2014_Catanzaro_book}.  Namely, for the model comparison we calculated a small set of models with slightly different surface gravity values ($\rm log g = [3.90, 3.95, 4.00]$), and different metallicity ($\rm M/H = [-0.3, -0.5] $), but the same effective temperature, $\rm T_{eff}= 9550 K$. The `Anders composition' was used for the abundances. 

For the search of the best stellar parameters, we used the flux calculations obtained using the standard `little' ODF with 1221 frequency bins \citep{Castelli_2005_DFSYNTHE}. The spectral flux observed for Vega was averaged on the same wavelength grid, and subsequently we broadened the fluxes using a Gaussian kernel with a full width at half maximum (FWHM) of 3nm below 350 nm, and with a FWHM of 5nm above 350 nm. 
The output of  MPS-ATLAS is the flux at top of the stellar atmosphere. To connect it to the observed flux one needs a scaling factor,  $\rm sf =  (d_{Vega}/R_{Vega})^2$, where $\rm R_{Vega}$ is Vega's radius and $\rm d_{Vega}$ is the distance of Vega to the observer. This factor is still uncertain and some assumptions must be made for the comparison \citep[see][where is was estimated to be $\rm sf  = (1.62\pm 0.07) \times  10^{16}$]{1994A&A...281..817C}. 

Here, we take this factor into account by taking the ratio of the observed flux to the modelled flux in the wavelength region 400 -- 600 nm, where the ratio showed the weakest dependence on the metallicity and surface gravity values. The scaling factor we obtained is $\rm sf = 1.59 \cdot 10^{16}$, which is within the uncertainty of the estimation. 
Overall, the best agreement of the modelled flux to the measured flux was achieved using $\rm M/H = -0.5$ and $\rm log g=3.90$, but very little difference is found between the flux for $\rm log g=3.90$ and the flux for larger surface gravity of $\rm log g= 4.00$.

For the comparison to PHOENIX, we downloaded the PHOENIX emergent intensities for $\rm T_{eff}=9400 K$, and  $\rm T_{eff}=9600 K$, and the surface gravity values $\rm log g = 3.50$ and $\rm log g = 4.00$, with the metallicity  $\rm M/H = -0.5$. After calculating the fluxes,  we had to interpolate to get to the effective temperature of $\rm T_{eff}=9550 K$,  and a surface gravity of $\rm log g = 3.90$. Due to the limited wavelength range for which the PHOENIX emergent intensities are provided we can not calculate the exact effective temperature from the flux.

The PHOENIX flux is averaged in exactly the same way as the IUE and STIS data, and subsequently all fluxes are broadened using a Gaussian kernel. 
Figure~\ref{fig:comp_model-Vega-STIS} shows the comparison in the UV  (120 nm -- 350 nm) between the measured flux, the flux calculated using the MPS-ATLAS code and the PHOENIX flux. 
The MPS-ATLAS calculation has an overall better agreement with the measurements than the PHOENIX modelled flux, but still deviates significantly for wavelength shorter than 260 nm.  Note that the elemental composition in the PHOENIX flux is the `Asplund-composition' by default, but scaled to $\rm M/H = -0.5$.

The comparison in the range 350 nm -- 1000 nm is displayed in  Fig.~\ref{fig:comp_model-Vega-STIS_part2}.  The  MPS-ATLAS calculations show very good agreement, with only a few spectral intervals with greater deviation than the measurement uncertainty. In contrast, there is an offset in the PHOENIX flux, whose deviation from the observations  shows a significant wavelength dependence that continues into the UV range (see Fig.~\ref{fig:comp_model-Vega-STIS}). While an overall shift could be explained by the scaling factor (which was taken to be the same as for MPS-ATLAS), the  wavelength dependence cannot be removed by a different normalisation. This indicates that the atmospheric structure or the continuum opacity has a slightly worse agreement. Moreover, the Paschen limit (820.4 nm) deviates significantly. The agreement of the modelled flux using MPS-ATLAS with the measurement is very good in the visible and near-infrared.

Finally, we compared calculated and observed  high-resolution spectra around hydrogen Balmer and Paschen lines. For that, we used the same atmospheric structure as before, but we calculated the high-resolution flux with resolving power $\rm R = 500 000$.  Subsequently, we used Gaussian broadening to degrade the resolving power to $\rm R = 400$. In Fig.~\ref{fig:Vega_Hlines} the observed and calculated flux together with the deviation of the calculation to the observations are shown.  The $\rm H_{\alpha}$ line shows the best agreement with at most 4\% deviations, while the deviations for the Paschen lines are up to 6\% and for higher Balmer lines even up to about 10\%. We note that in a previous comparison \citep{Bohlin_Vega_2004}  a better agreement (with deviations of a few percent) was achieved. However, these differences are mainly attributed to differences in the line depth of the updated STIS data (see Figure 4 in \citet{Bohlin_Vega_2004}), while some small deviations might be a result of modifications and updates in the MPS-ATLAS code compared to the original ATLAS9 at that time.


\section{Summary and conclusion}
\label{Sec:Discussion}
We presented the structure and extended functionality of the MPS-ATLAS code. The code is based on the ATLAS9 code, but has been extended to allow flexible and faster handling of ODF calculations, as well as emergent flux calculations. This makes the code suitable for radiative transfer calculations along rays from 3D MHD cubes. 
Furthermore, the atmosphere model calculations were sped up and made more user-friendly. 
We  also improved the equilibrium number density calculations, and included NH photo dissociation opacity. The code-to-observations comparison showed that MPS-ATLAS gives excellent agreement with the  observed  solar spectrum and Vega spectrum (better than some other widely used codes). \\

The source code is available on request along with a set of testing input files. A detailed explanation of the input files is given in Appendix~\ref{app:example_input}. Furthermore, an online tool for calculating ODFs, stellar models and fluxes will soon be released. 

A fine grid of stellar models, fluxes and CLVs is being calculated (Kostogryz, et al. in prep.).  
This grid will cover the range of effective temperature between 3500K and 9000K in 100K steps, a range of surface gravity from $\rm logg = 3.0$ to $\rm logg = 5.0$, and  metallcities between -5.0 to 1.5 using very fine steps around solar metallicity of 0.05 dex.

\begin{acknowledgements}
This work has received funding from the European Research Council (ERC) under the European Union's Horizon 2020 research and innovation programme (grant agreement No. 715947). This work has been partially supported by the BK21 plus programme through the National Research Foundation (NRF) funded by the Ministry of Education of Korea and by the Max Planck Society grant "PLATO Science" and DLR PLATO grant Nr. $50$OO$1501$ and $50$OP$1902$. YCU acknowledges funding through STFC consolidated grants ST/S000372/1.

This work has made use of the VALD database, operated at Uppsala University, the Institute of Astronomy RAS in Moscow, and the University of Vienna.
\end{acknowledgements}

\bibliographystyle{aa} %
\bibliography{bib} %

\newpage
\begin{appendix}

\section{Solving the statistical equilibrium}
\label{app:solve_Enumbers}
 
MPS-ATLAS assumes local thermodynamic equilibrium (LTE), which is a good enough approximation for photospheres in cool main-sequence stars. 
In LTE, the populations of atomic and molecular levels  are in thermal equilibrium and, thus, can be evaluated independently of the  the radiation field with the help of the Saha-Boltzmann (SB) and Guldberg-Waage equations \citep[with the latter being the analogue of the Saha-Boltzmann equation for molecular chemical equilibrium calculations, see, e.g.,][]{Tatum1966}.  

Following ATLAS9, in the MPS-ATLAS code, a set of equilibrium equations for all species together with conservation constraints is formulated. Then, using the SB equation, the set of equations are expressed in terms of neutral atom  and electron number densities. 
For calculations without molecules, the species taken into account in the set of equilibrium equations are hard-coded and include the most important electron donors, H, He, C, Na, Mg, Al, Si, K, Ca and Fe. 
For calculations with molecules, the number of species, $\rm nmol$,  is set by the number of elements and molecules listed in the file `molecules.dat', whose default version includes the atoms listed above and significantly more. In this file, not only the species are listed, but also the molecular equilibrium constants are provided together with their dissociation energies, $\rm D_0$. The equilibrium constants are pre-tabulated as a function of temperature using coefficients from fits to the NIST-JANAF tables \citep{JANAF_04}, and the dissociation energies were taken from \citet{Huber1979}. These pre-tabulated coefficients are used to evaluate the Guldberg-Waage equation. 

For both cases, the equilibrium equations  can be rearranged, such that for each considered species, i, one has an equation $ {f}_i (n_1, ..., n_i, n_{total}) = 0 $, where $n_i$ is the number density of the species i. To find the number densities, the matrix $M_{ij} = \partial f_{i}/ \partial n_j$ is set up, and the matrix equation 
\begin{equation}
    \vec{M} \cdot \vec{\Delta n} = \vec{f},
\end{equation}
where $\vec{\Delta n}$ is the change in $\vec{n}$, is solved using the method of triangular decomposition \citep[Chapter 9]{ralston1978first}. Using a Newton-Raphson technique, the solutions are iterated until the relative change for each species in number density, $\Delta n_i/n_i$, becomes less than $10^{-4}$. A more detailed description of the underlying procedure can be found in \citet{Kurucz_manual_1970}.  

\section{Improvements for model calculations}
\label{app:Improv_model}

Since calculating atmosphere models is an iterative process, it is useful to know when to stop. While in ATLAS9 there were no criteria that identified when the atmosphere model is close enough to RE, we implemented the following procedure. 
For the model convergence criteria the greatest relative temperature adjustment in all atmosphere points has to be smaller than $10^{-5}$. If this criterion is reached before the prescribed maximum number of iterations, it is considered that the atmosphere model has converged. This threshold can be easily changed.  For reference, the greatest relative temperature adjustment is written out in the last iteration. Moreover, atmosphere models are calculated on a prescribed Rosseland mean optical-depth, $\rm \tau_{Ross}$, grid. We improved the setup to make it more user friendly  (for an example setting see Appendix~\ref{app:example_input}). %
The treatment of convective flux and overshoot in the code is mainly kept as originally implemented (for more details see \citet{Castelli_1996_convection}).  The so-called `approximate overshoot' was tested in \citet{Castelli_1997_mix_os}.

\section{Starting model for RE calculations}
\label{app:Re-calc}

For the model calculations, the initial 1D atmosphere model is re-scaled using the ratio of the desired effective temperature value to its  initial value on a Rosseland mean optical-depth, $\tau_{\rm Ross}$ grid.  This is achieved using the temperature as a function of  $\tau_{\rm Ross}$ and applying
\begin{equation}
 \rm   T(\tau_{Ross}) = \frac{T_{eff}}{T_{eff}^{\rm initial}} T(\tau_{Ross})^{\rm initial},
\end{equation}
where the superscript `initial' indicates the initial model temperature structure, and its effective temperature. \citet[Chapter 2.12]{Kurucz_manual_1970} showed that the Rosseland optical depth is suitable for such a re-scaling. The re-scaled temperature structure is converted back onto the column mass grid. This results in a first starting point model for which subsequently the RTE has to be solved. 


\section{Solving Radiative Transfer}
\label{app:RT-calc}
%

\begin{figure}
  \centering 
  {\includegraphics[width=1.0\linewidth]{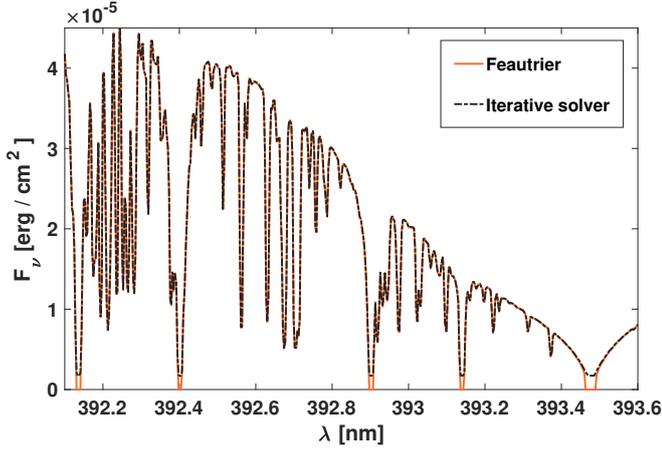}}
  \caption{High-resolution disc-integrated flux, $F_{\nu}$, calculated using the iterative solver and the Feautrier solver with 3 view angles. }
  \label{fig:f_i_hr}
\end{figure}

\begin{figure}
  \centering 
  {\includegraphics[width=1.0\linewidth]{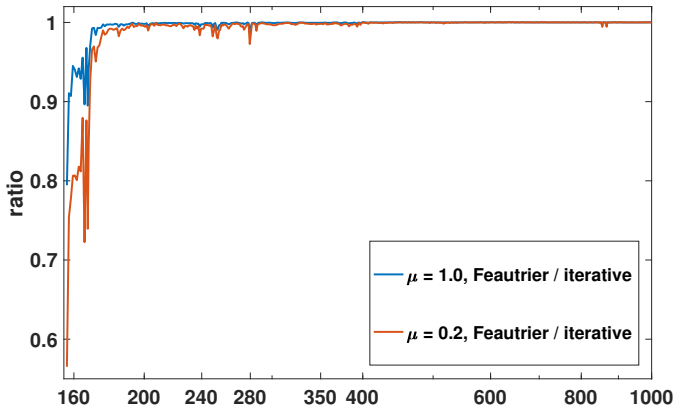}}
  \caption{Ratio of the emergent intensities for two different limb positions with the ODF approach calculated using the Feautrier solver to the one using the iterative solver.}
  \label{fig:f_i_odf}
\end{figure}

The MPS-ATLAS code has two different radiative solvers implemented. In the original ATLAS9 implementation of the RTE solver by R.~L.~Kurucz, the scattering part of the source function is found iteratively (hereafter, iterative solver).  To minimise the computational time,  pre-tabulated  matrices for the evaluation of the mean intensity on a fixed optical depth grid are used \citep[for more details see][]{Kurucz_1969ApJ, Kurucz_manual_1970}.
The alternative way to account for the scattering part of the source function is to use a Feautrier method as described in \citet{FRH_Mihalas} (hereafter, Feautrier solver). In this approach, the second-order transfer equation, derived using Feautrier variables together with upper and lower boundary conditions, is solved. The Feautrier method allows to set the number of viewing angles, $\mu$,  which are included in the calculation,  to three, four or eight \citep{Lester_Neilson2008}. 
The two solvers have a different treatment of the source function at frequencies where strong lines are formed in the upper atmosphere, i.e.the case when the top atmospheric point has $\rm \tau_{\nu} > 0.2$. Since for such frequencies the atmosphere is not sufficiently high in order to obtain an accurate solution of the RT and non-LTE effects are important, the Feautrier solver sets the source function and the flux to zero (see orange line in Fig.~\ref{fig:f_i_hr}). This differs from the  iterative method that still solves the RT equation in such cases, but makes an approximation by considering a prolongation of the atmosphere with a source function, $\bar{S}$, equal to the $\bar{S}(\rm top)$  value. An example calculation of the high-resolution flux for a small wavelength interval obtained using the two different solvers is shown in  Fig.~\ref{fig:f_i_hr}. The two solutions are both approximations, but lead to different emergent intensities, as can be seen by looking at the Ca II K line ($\lambda =393.478$ nm).  Note, that the different treatment has a negligible effect when ODFs are used for wavelengths longward of 200 nm  (see Fig.~\ref{fig:f_i_odf}).

We tested the performance of these two solvers. For that we compare the resulting effective temperature calculated from the emergent flux with the value set in the atmosphere model calculation, and we measure the computation time needed to reach a converged model. We found that, generally, the iterative solver is the most accurate but slower compared to the Feautrier solver. The optimal trade-off between computational time and accuracy is achieved when using the Feautrier solver with three $\mu$ angles.
Ideally, to obtain consistent results the same solver should be used  for  the specific emergent intensities  as is used for the radiative equilibrium calculations. 
 
\section{ODF and opacity calculations} 
\label{app:ODF-impl}
The calculation and pre-tabulation of ODFs on a T-P grid is done in three steps. First, the  quantities needed for selecting which atomic and molecular lines to include and for calculating the line opacity are computed (e.g., molecular and atomic number densities and continuum opacity). Second, lines are selected and the high-resolution opacity is calculated. Finally, the high-resolution opacity is split into bins, sorted and averaged over the sub-bins.

In the first stage, the T-P grid together with the element composition is set up. Subsequently, in each T-P point the equilibrium number densities are found as described  in Sect.~\ref{app:solve_Enumbers}. Having this information the following quantities are obtained: i) the number density over partition function, $\rm N_j / U_j$, where $\rm j$ indicates the ionisation stage, are obtained up to $\rm j=5$ for all atoms, ii) the total continuous opacity, $\kappa_{c}$, on a grid of 858 frequency points in the range from \mbox{1 nm -- 500000 nm}, iii)  the quantity $\sqrt{\rm  2k_{B} T/ \rm  m_{el}} / c$, which is the ratio of the thermal velocity of a given element (with atomic mass $\rm m_{el}$) to the speed of light, for micro-turbulence $\rm v_{turb} = 0\, \rm km/s$, and iv) the quantity 
\begin{equation*}
\rm nf_{max}   = \left( \frac{1}{\rho \Delta v_D } \frac{\rm N_j (el)}{\rm U_j (el)}\right) \bigg/ \left(\rm cutoff \cdot  \rm min(\kappa_c ) \right)  , 
\end{equation*}
which gives an estimated  measure of the line opacity relative to the minimum continuous opacity $\rm min(\kappa_c)$ in a given frequency bin multiplied by a cutoff, where the cutoff typically has a value of order $10^{-3}$.  In this ratio $\rho$ is the density, 
and  the Doppler shift of the transition at the frequency $v_0$ is 
\begin{equation}
     \Delta v_D = \frac{v_0}{c} \sqrt{\frac{\rm 2 k_B T}{\rm m_{el}} + \rm v_{turb}^2}. 
\end{equation}

These quantities are required in the second stage for two purposes: I) for the pre-selection of lines in order to reduce the computational cost, where only lines are selected that pass several conditions.  The first condition is that the quantity $\rm nf_{max}$ has to be greater than unity, which ensures that without the knowledge of the oscillator strength there are enough particles in a given ionisation stage. 
For the final condition, the oscillator  strength, $\rm f_{ij}$, of the transition,  the statistical weight of the i-th level, $\rm g_i$, and the Boltzmann factor are taken into account.  
This results in 
\begin{equation*}
    \frac{\sqrt{\pi}e^2}{m_e c} g_i f_{ij} e^{\left(-\chi_i / k_B T\right) } \cdot \rm nf_{max} > 1.
\end{equation*}
This conditions ensures that the line core opacity is greater than a thousandth of the continuous opacity.


II) the pre-calculated quantities are needed for calculating  the line-strength and broadening, as they depend on the ionisation fraction and the Doppler shift. Namely, the line absorption coefficient as given in Eq.~\eqref{eq:l_nu_line}
is computed. Here, for all lines, except hydrogen lines, the Voigt-Profile is used.  For hydrogen lines the Stark broadened profile is used \citep{SYNTHE_2002_Cowley}. With the pre-selected lines and the line-profiles, the detailed high-resolution opacity is calculated on a wavelength grid from 8.9766 nm to 10 000 nm  with a resolving power of $\rm R = 500 000$. This results in the wavelength points
\begin{equation}
  \rm   \lambda_{n} = 8.9766 \left(1.0 + \frac{1.0}{R}  \right)^{n-1} \,  nm,
\end{equation}
where $\rm n$ is the index of the grid points.

During the third stage the high-resolution opacity is split into wavelength bins, which are either on the standard Kurucz's grid, or user-defined. The standard wavelength grid can be set by the keyword `binsizes on' and is hardcoded. For a user-defined bin grid, an additional file, `bin-grid-sizes.dat' that contains the bin borders, has to be provided. After selecting the bin range, the opacity in each bin is sorted and further split into sub-bins as described in Sect.~\ref{subs:ODF_approach}. The number of sub-bins and their sizes are either as in the standard Kurucz's configuration, which is automatically set if the standard bin grid  is used, or a user-defined configuration, which has to be specified in the file `subbin-info.dat'.  Finally, the opacity is averaged over the sub-bins, and written out.

Having  generated a ODF table, the sub-bin opacity values  can be read, and further processed in module II and module III. In both modules, the sub-bin opacity is added to the continuum opacity  for each sub-bin separately. Subsequently, the  RTE is solved. The resulting quantities, such as the flux $H_{\nu}$, and its derivative in module II, or the surface flux or surface intensity, $I_{\nu}$, in module III, need to be calculated for the frequency bin. Thus, an average over the weighted contributions from the sub-bins is used. The weights correspond to the wavelength interval of the sub-bins. As an example, the surface intensity in a bin i, $I_{bin,i}$, with $\Delta \lambda_i$ that is  centred at $ \lambda_c = (\lambda_{i+1} +\lambda_i) \cdot 0.5$, is obtained by summing the calculated intensity in each sub-bin, weighted by the sub-bin width, $w_s$, as follows
\begin{equation}
    I_{bin,i} =  \sum_{s=1}^{s= n_s} w_{s,i} I_{\lambda_c}(\kappa_{tot, s,i}),
\end{equation}
where $I_{\lambda_c}$ is the intensity at $\lambda_c$ obtained using the  capacities along the atmosphere,   $\kappa_{tot, s,i}$, which are  the sum of the continuous opacity at $\lambda_c$ and the averaged sub-bin opacity of the sub-bin s in each atmosphere point.

\section{Molecular photo-dissociation}
\label{app:molec_opacitiy}
ATLAS9 includes photo-dissociation for CH and OH included.  This is achieved by using pre-tabulated cross-sections and  corresponding partition functions. 
In addition, we implemented the NH dissociation opacity. Our implementation is, however, slightly different from that of CH and OH.
%
The number densities for NH that are obtained from the equilibrium calculations are multiplied with the cross-sections directly. Since a negligible fraction of NH might be 
in an excited state, this treatment potentially overestimates the opacity. 

\begin{figure}
  \centering 
  {\includegraphics[width=1.0\linewidth]{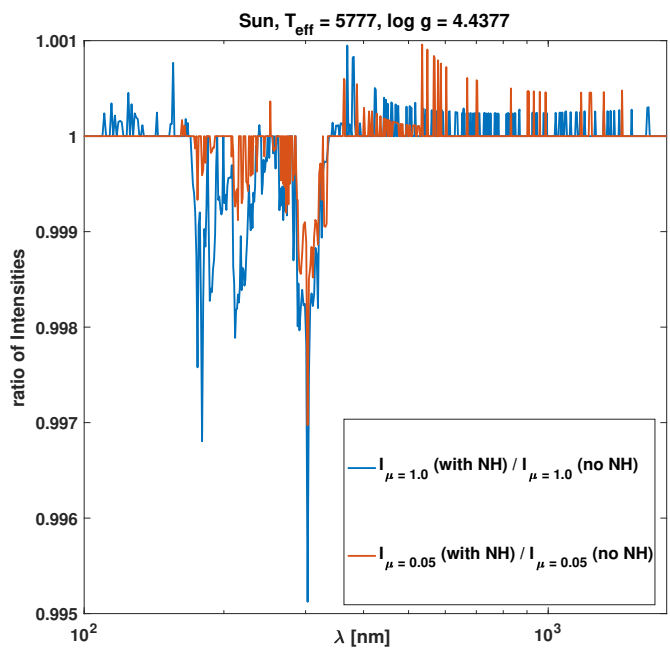}}
  \caption{ Ratio between emergent intensities for the Sun obtained including NH photo-dissociation and without it for two different viewing angle $\mu$.}
  \label{fig:w-no-NH}
\end{figure}

To test the effect of NH photo-dissociation opacity on the specific intensities,  we calculated the solar atmosphere model in RE with and without the NH photo-dissociation, and subsequently the corresponding emergent intensities. The ratio of the emergent intensities is shown in Fig.~\ref{fig:w-no-NH}. It indicates that the difference is below 0.5\% at disc centre and below 0.3\% at the limb ($\mu = 0.05$). Thus we conclude that including NH is not sufficient to account for the missing opacity in the UV.    

\section{Example input}
\label{app:example_input}

\begin{figure*}
\begin{verbatim}
! CONTROL file for MPS-ATLAS
calc_odf off
calc_flux
calc_model off

odf_input =./INPUT/odf.input
odf_tp_grid =./INPUT/odf.tpgrid
odf_output =./INPUT/ODF.nc


model_input=./INPUT/model.input
model_start= ./INPUT/model.start
model_odf=./INPUT/ODF.nc

flux_input =./INPUT/flux.input
flux_model = ./INPUT/flux.model
flux_odf =./INPUT/ODF.nc
endfile
 \end{verbatim}
     \caption{Example of an `mpsa.control'   file.}
    \label{fig:atlas_control}
\end{figure*}

To run MPS-ATLAS several input files are required. In an overall control file, `mpsa.control', (shown in Fig.~\ref{fig:atlas_control}) the user has to specify which module or modules should be executed and give the names of the input files. To set a module for calculations a line with the keyword `calc\_' followed by either `odf', `model' or `flux' has to be included. By adding an `off' after the keyword the module can be switched off again without deleting the line. In the example in Fig.~\ref{fig:atlas_control} the code will only calculate the emergent flux module.

For every module there are three kind of input files that needs to be specified: i) the input file that contains computational flags, ii) a model atmosphere file, and iii) the file that contains the opacity tables, where for the ODF table calculation the name of the output file is specified. 
The order of the lines in the control file is irrelevant, but to indicate that there are no more lines to read the keyword `endfile' has to be given.

\subsection{Input files}

To start a particular module, it has to be activated in the `mpsa.control' file. Subsequently, each module requires different settings in the `.input' file, where some settings are common. The order of the keyword lines in the input file is not important, except for the last two lines which have to be `\underline{begin}' and `\underline{end}', to indicate the end of the input file and start the calculation.  Example input files for each module are given in  Fig.~\ref{fig:input00}--Fig.~\ref{fig:input02}. 
All three modules have most of the flags in common, while a few are specific for certain modules.\\

\subsubsection{Common settings}
In any of the modules, molecules can be included or excluded by the keywords `\underline{molecules on}' or `\underline{molecules off}'. If the molecules are `on', a routine reads the file `molecules.dat' which should be located in the INPUT folder. \\

For all modules the chemical composition needs to be specified (as shown in Fig.~\ref{fig:input00} -- Fig.~\ref{fig:input02}. The line starting with the keyword `\underline{abundance scale}' specifies the metallicity. The number following this keyword scales all elements heavier than helium. Consequently, a single scaling factor accounts for changes in metallicity, so that  it is not necessary to change all elements individually when changing  the metallicity. All element abundances can be set using the keyword `\underline{abundance change}' followed by the elemental number and the number fraction. While for hydrogen and helium the number faction is given as   $\rm N_{element} / N_{total}$, for all other elements it is given in $\rm \log10(N_{element} / N_{total})$.  For consistency  the  settings in all input files should be kept the same throughout the calculation of a given star.\\

For the  opacity calculations the continuous opacity sources as listed in Sect.~\ref{subsection:calc_the_op} are all included by default. If any of them need to be excluded they can be switched off by adding `\underline{opacity off}' followed by keyword of the corresponding opacity source. The continuous opacity sources are grouped as follows: 
\textbf{1)} $\text{H I}$ bound-free transitions (bf) and free-free transitions (ff), keyword `H1' 
\textbf{2)} $\text{H}_{2}^{+}$ bf and ff, keyword `H2+'
\textbf{3)} $\text{H}^{-}$ bf and ff, keyword `H-'
\textbf{4)} Rayleigh scattering on $\text{H I}$, keyword `Hray' 
\textbf{5)} $\text{He I}$ bf and ff, keyword `He1'  
\textbf{6)} $\text{He II}$ bf and ff,  keyword `He2'
\textbf{7)} $\text{He}^{-}$  ff, keyword `He-'
\textbf{8)} Rayleigh scattering on $\text{He}$, keyword `Heray'
\textbf{9) -- 11)} various bf and ff transitions of C, N, O, Ne, Mg, Al, Si, Ca, Fe, the molecules CH, OH and NH, and their ions, keywords `Cool', `Luke', and `Hot'
\textbf{12)} electron scattering, keyword `Elec'
\textbf{13)} Rayleigh scattering on $\text{H}_2$, keyword `H2ray'.
In addition, if the line opacity should be excluded in either the model calculations or the flux calculations this can be achieved by including the line `\underline{opacity off Lines}'.  \\

The keywords `\underline{user defined binning}' switches between the original standard bin and sub-bin configuration used in \citet{Castelli_2005_DFSYNTHE} if set `off', and a user-defined configuration if set `on'. Having set a user defined binning and sub-binning in the input files, individual grids have to be provided in separate input files, where the number of grid points, their intervals, and the number of sub-bins, and their sizes are specified. Then, the code will allocate the requested number of bins and sub-bin. This switch sets the bin and sub-bin configuration for module I, but has to be consistent with the format of the ODF used in module II and III.\\

The two keywords `\underline{print}' and `\underline{punch}' control how much information is written out during a run. They are both followed by an integer. For `\underline{print}' the following values can be set: 
\begin{itemize}
    \item 0 for no print
    \item 1 for summary tables
    \item 2 prints the temperature corrections, and surfaces fluxes
    \item 3 prints everything listed as for point 2, and also $\tau_{\nu}$, $S_{\nu}$ and $J_{\nu}$ for each frequency
    \item 4 prints  the quantities listed under point 2 and all opacities for each frequency
\end{itemize}
For `\underline{punch}' there are only four options:
\begin{itemize}
    \item 0 nothing is written out
    \item 1 writes atmosphere model 
    \item 2 writes punch 1 and in addition surface fluxes or intensities 
    \item 5 writes punch 2 and  also molecular number densities over partition functions
\end{itemize}

\subsubsection{ODF generation specific settings}

A specific switch for the ODF calculations can turn filter calculations on or off by using the keyword `\underline{filter}' \citep[for more details on filter ODF see][]{Anusha_2020}. If this switch is not set, the filter calculations are switched off by default. \\ 

Module I has additional keywords: In the default setting at least one ODF table using  the micro-turbulence velocity of $\rm v_{turb} = 0.0\, km/s$ is calculated. To obtain additional ODF tables the keyword `\underline{velocities values}', followed by an integer indicating   how many  turbulent velocity values should be calculated, can be set. After the integer, the code expects the corresponding number of float numbers setting the micro-turbulence velocities. Moreover, by default the opacities in the sub-bin are averaged  using the geometric average. This can be changed to an arithmetic average by setting the keyword combination `\underline{mean arithmetic'}.   \\

\begin{figure*}
    \centering
\begin{verbatim}
molecules on
print 2
punch 2
velocities values 3 0.0 1.5 2.0
odf number 5
user defined binning off
odf mean arithmetic 
filter off
abundance scale   1.9498 abundance change 1 0.91100 2 0.08900
 abundance change  3 -10.88  4 -10.89  5  -9.44  6  -3.48  7  -3.99  8  -3.11
 abundance change  9  -7.48 10  -3.95 11  -5.71 12  -4.46 13  -5.57 14  -4.49
....
 abundance change 87 -20.00 88 -20.00 89 -20.00 90 -12.02 91 -20.00 92 -12.58
 abundance change 93 -20.00 94 -20.00 95 -20.00 96 -20.00 97 -20.00 98 -20.00
 abundance change 99 -20.00
begin
end
\end{verbatim}
    \caption{Example input for ODF generation (odf.input).}
    \label{fig:input00}
\end{figure*}

\begin{figure*}
    \centering
\begin{verbatim}
molecules on
iterations 15
overshoot on
print 4
punch 2
user defined binning on
wave limits 120.0  10000.0 
convection on 1.25
turbulence off 0.0 0.0 0.0 0.0
starting model teff 4200.0 surface gravity 4.200
scale b -6.800 0.100 2.000 4000 4.800
abundance scale   1.9498 abundance change 1 0.91100 2 0.08900
 abundance change  3 -10.88  4 -10.89  5  -9.44  6  -3.48  7  -3.99  8  -3.11
 abundance change  9  -7.48 10  -3.95 11  -5.71 12  -4.46 13  -5.57 14  -4.49
......
begin
end
\end{verbatim}
    \caption{Example input file for model calculations (model.input)}
    \label{fig:input01}
\end{figure*}

\begin{figure*}
    \centering
\begin{verbatim}
molecules on
recalxne on
user defined binning off
frequencies little
wave limit 100.0 2000.0
surface intensity  11 1.0 0.9 0.8 0.7 0.6 0.5 0.4 0.3 0.2 0.1 0.05
print 0
punch 2
abundance scale   1.9498 abundance change 1 0.91100 2 0.08900
 abundance change  3 -10.88  4 -10.89  5  -9.44  6  -3.48  7  -3.99  8  -3.11
 abundance change  9  -7.48 10  -3.95 11  -5.71 12  -4.46 13  -5.57 14  -4.49
......
begin
end
\end{verbatim}
    \caption{Example input for specific emergent intensities (flux.input).}
    \label{fig:input02}
\end{figure*}

\begin{figure*}
    \centering
\begin{verbatim}
          1
          11
 1.7781904E-04   2469.5  1.778E+01  2.186E+08  8.913E-04  1.911E-02  1.500E+05
 2.2044887E-04   2482.8  2.204E+01  2.674E+08  1.034E-03  1.969E-02  1.500E+05
 2.6742791E-04   2496.6  2.674E+01  3.217E+08  1.161E-03  2.024E-02  1.500E+05
 3.2072459E-04   2510.9  3.207E+01  3.837E+08  1.279E-03  2.085E-02  1.500E+05
 3.7984064E-04   2517.4  3.798E+01  4.461E+08  1.491E-03  2.112E-02  1.500E+05
 4.4418566E-04   2523.7  4.442E+01  5.131E+08  1.712E-03  2.138E-02  1.500E+05
 5.1520221E-04   2530.2  5.152E+01  5.865E+08  1.941E-03  2.167E-02  1.500E+05
 5.9445852E-04   2537.0  5.945E+01  6.679E+08  2.179E-03  2.198E-02  1.500E+05
 6.8367171E-04   2544.2  6.837E+01  7.592E+08  2.428E-03  2.232E-02  1.500E+05
 7.8480058E-04   2551.9  7.848E+01  8.625E+08  2.689E-03  2.270E-02  1.500E+05
 9.0013518E-04   2560.2  9.001E+01  9.802E+08  2.960E-03  2.311E-02  1.500E+05

\end{verbatim}
    \caption{Example of model atmosphere (model.start or flux.model).}
    \label{fig:model}
\end{figure*}

\subsubsection{Model calculation specific settings}

For the atmosphere model calculations, the desired stellar parameters, and the Rosseland grid need to be specified. This is controlled by using the keyword `\underline{scale}' which has three different options: 
\begin{itemize}
    \item  no letter follows the `\underline{scale}' keyword; the code reads in the number of depth points of the model, the starting Rosseland mean optical-depth, $\tau_{\rm Ross}$, in log10, the step-length in log10, the effective temperature, and the surface gravity, $\rm log g$;
    \item if the letter `\underline{x}' follows, instead of the  step-length, the maximum $ \tau_{\rm Ross}$ is read
    \item if the letter `\underline{b}' follows, instead of the number of depth points, the starting  $ \tau_{\rm Ross}$, the step-length, and the maximum $\tau_{\rm Ross}$ is read, before $\rm T_{eff}$, and log g
\end{itemize}
The keyword `\underline{iterations}' has to be set for the model calculations indicating the maximal number of iterations to be executed. Note, that the calculations will be stopped if the model converges before this number is reached. \\
For this module the effective temperature and surface gravity of the starting model, which is used for the initial re-scaling (see Sect.~\ref{app:Re-calc}), need to be specified. Thus, a line starting with the keywords `\underline{starting model}' followed by the `\underline{teff}' and `\underline{surface gravity}' values has to be included.\\

By default the convective flux and overshoot is turn off. To change that, the keyword `\underline{convection}' sets the convection flux calculations `on' or `off', and a line with `\underline{overshoot on}' turns on the additional convection flux in the overshoot region \citep{Castelli_1996_convection}. The mixing-length value is the number following  `\underline{convection on}', and is usually set between 0.0 and 2.0. Moreover,  turbulence can be set on/off, by including the line `\underline{turbulence}' followed by four numbers. The turbulent velocity, $\rm v_{turb}$, is computed using the four numbers a, b, c and d after `\underline{turbulence on}', as follows:  
\begin{equation}
   \rm v_{turb} = a \, \rho^{b} +c \, v_{snd}/(1.0^{-5}) +d,
\end{equation}
where $\rm v_{snd}$ is the sound speed in centimetres per second, and the term $d$ accounts for the convective turbulence, which is often the only contribution used on late-type stars. \\

\subsubsection{Settings for model and flux calculations}
For module II and module III in the case of two standard frequency grids, the keyword `\underline{frequencies}'  determines if the `big' or `little' Kurucz grid is used. Moreover, if for the calculation only a particular wavelength range should be considered, the keyword `\underline{wave limits}' followed by the starting and end wavelength in nanometre  can be specified. If this keyword is not set, the whole available wavelength range of the considered opacity table will be used.

\subsubsection{Flux calculation specific settings}
Module III has a switch to recalculate the electron number densities in a given atmosphere using the given elemental composition. This switch is in particular important if  atmospheres from a 3D MHD cube where $\rm n_{e}$ are not pre-calculated are used;  it can be switched on using `\underline{recalxne on}'.
Finally, setting the keyword `\underline{surface flux}' starts the calculations for the emergent flux at the surface, whereas, the keyword `\underline{surface intensity}', followed by the number of view angles and their values, results in emergent intensity calculations at the specified view angles.

\subsection{Model structure files}

For all modules a second input file is read. For the ODF table calculations (module I), this file contains the T-P grid on which the ODF table should be calculated, while for modules II and III, it contains the starting atmosphere model and the atmosphere model for which the emergent intensities are calculated, respectively. The structure of the T-P grid (odf.tpgrid) is simple, the first two lines are two integers indicating the number of T and P points. They are followed by the temperature values given in each line, and  the pressure values. 

The atmosphere model for modules II and III has the structure shown in Fig.~\ref{fig:model}. The first line gives the number of models, the second line lists the number of depth points. The first line is always 1, except if models form 3D MHD cubes are calculated. Starting in the 3rd line the model is given for each depth point using seven columns. The first column lists the column mass, followed by the temperature, pressure, electron number density, mean Rosseland opacity, radiation pressure, and turbulent velocity in each depth point. We only consider cases with constant turbulent velocity, while the code allows to use a depth dependent $\rm v_{turb} $, when several ODF tables for a range of   $\rm v_{turb} $ are read in.

\end{appendix}

\end{document}